\documentclass[aps,pra,reprint,onecolumn,superscriptaddress,showpacs,nofootinbib]{revtex4-1}
\usepackage{graphicx,lipsum}
\usepackage{amsmath,amssymb,color}
\usepackage{pgfplots}
\usepackage[export]{adjustbox}
\usepackage[normalem]{ulem}
\usepackage{bm}

\newcommand{\bolds}{{\bf{s}}}

\newcommand{\boldv}{{\bf{v}}}
\def\refeq#1{(\ref{#1})}

\newcommand{\bom}{{\mbox{\boldmath $\omega$}}}
\newcommand{\bomhat}{{\mbox{\boldmath ${\hat \omega}$}}}

\definecolor{violet1}{RGB}{153 0 204}
\definecolor{dark_green2}{RGB}{0 102 0}

\def\red#1{\textcolor{black}{#1}}
\def\blue#1{\textcolor{black}{#1}}

\def\violet#1{\textcolor{black}{#1}}

\def\dgreen#1{\textcolor{black}{#1}}

\begin{document}
\title{Dissipation anomaly in a turbulent quantum fluid}
\author{Luca~Galantucci}
\affiliation{Istituto per le Applicazioni del Calcolo `M. Picone’,
IAC-CNR, Via dei Taurini 19, 00185 Roma, Italy}

\affiliation{Joint Quantum Centre (JQC) Durham--Newcastle, 
School of Mathematics, Statistics and Physics, 
Newcastle University, Newcastle upon Tyne, 
NE1 7RU, United Kingdom}
\email{l.galantucci@iac.cnr.it}
\author{Em~Rickinson}
\affiliation{Joint Quantum Centre (JQC) Durham--Newcastle, 
School of Mathematics, Statistics and Physics, 
Newcastle University, Newcastle upon Tyne, 
NE1 7RU, United Kingdom}
\author{Andrew~W.~Baggaley}
\affiliation{Joint Quantum Centre (JQC) Durham--Newcastle, 
School of Mathematics, Statistics and Physics, 
Newcastle University, Newcastle upon Tyne, 
NE1 7RU, United Kingdom}
\author{Nick~G.~Parker}
\affiliation{Joint Quantum Centre (JQC) Durham--Newcastle, 
School of Mathematics, Statistics and Physics, 
Newcastle University, Newcastle upon Tyne, 
NE1 7RU, United Kingdom}
\author{Carlo~F. Barenghi}
\affiliation{Joint Quantum Centre (JQC) Durham--Newcastle, 
School of Mathematics, Statistics and Physics, 
Newcastle University, Newcastle upon Tyne, 
NE1 7RU, United Kingdom}

\affiliation{Joint Quantum Centre (JQC) Durham--Newcastle, 
School of Mathematics, Statistics and Physics, 
Newcastle University, Newcastle upon Tyne, 
NE1 7RU, United Kingdom}
\date{today}

\begin{abstract}
When the intensity of turbulence is increased (by increasing the 
Reynolds number, \textit{e.g.} by reducing the viscosity of the fluid), 
the rate of the dissipation of kinetic energy decreases but does not tend 
asymptotically to zero:
it levels off to a non-zero constant as smaller and smaller vortical
flow structures are generated.  This fundamental property, 
called the {\it dissipation anomaly}, is sometimes referred to as the 
zeroth law of turbulence.  The question of what happens in the limit 
of vanishing viscosity (purely hypothetical in classical fluids)
acquires a particular physical significance 
in the context of liquid helium, a quantum fluid which becomes 
effectively inviscid 
at low temperatures achievable in the laboratory.  
By performing numerical simulations and  
identifying the superfluid Reynolds number, here we show evidence 
for a superfluid analog to the classical dissipation anomaly. 
Our numerics indeed show that as the superfluid Reynolds number increases,
smaller and smaller structures are generated on the quantized vortex lines 
on which the superfluid vorticity is confined, balancing
the effect of weaker and weaker dissipation.
\end{abstract}
\maketitle

\section{Introduction}

It is well known from experiments and numerical simulations of
incompressible, homogeneous and isotropic turbulence that, if the
fluid's kinematic viscosity $\nu$ tends to zero (or, equivalently,
if the Reynolds number tends to infinity), the
average dissipation rate of turbulent kinetic energy does not decrease to zero,
but tends to a finite constant \cite{Sreeni1984,KanedaIshihara2003}. 
In other words, the limit of
the \red{incompressible}
 Navier-Stokes equation for vanishing viscosity is not the
Euler equation, as one would naively expect. \blue{This}
\textit{dissipation anomaly}, led Onsager \cite{Onsager1949,EyinkSreeni2006}
to conjecture that the
solution of the Euler equation is not a smooth velocity field:
smaller viscosities are compensated by the creation of motions
at smaller and smaller length scales containing much vorticity 
but little energy. The dissipation anomaly is thus related to
the properties of turbulence at the smallest length scales of the flow.

Progress in low temperature physics adds a twist to this
story. Turbulence with vanishing viscosity, in fact, is not a
mathematical idealisation but can be created
in the laboratory by cooling liquid helium (either isotope $^4$He or $^3$He)
\blue{below the critical temperature for Bose-Einstein condensation.}
Below this temperature, liquid helium becomes a quantum fluid 
consisting of two interpenetrating components: the inviscid \textit{superfluid}
(associated to the quantum ground state) and a gas of thermal
excitations (the \textit{normal fluid}) which carries 
\blue{entropy and viscosity}.  Upon further cooling, the amount 
of thermal excitations decreases rapidly; for example, 
$^4$He becomes effectively a pure superfluid below $1~\rm K$. 
Turbulence is easily generated in this superfluid component by 
mechanical or thermal stirring, 
and consists of a disordered tangle of vortex lines  
\blue{of quantised circulation.} 

\blue{Experiments and theory} have revealed
that, despite the two-fluid nature and the quantised circulation, 
in certain regimes and at length scales larger than the average 
inter-vortex spacing $\ell$, superfluid turbulence may show properties
similar to ordinary (classical) turbulence 
\cite{BLR2014,stalp-skrbek-donnelly-1999}. A notable example is the 
observation in liquid helium \cite{maurer-tabeling-1998,salort-etal-2010} 
of the famed Kolmogorov energy spectrum \cite{kolmogorov-1941}
revealing \blue{an energy cascade} at those large length scales.
The aim of this letter is to present \blue{evidence}
of an additional similarity between classical and quantum
turbulence, this time occurring at the smallest length scales of the flow: 
\blue{a superfluid analog of} the classical dissipation anomaly.
After \blue{defining} the superfluid Reynolds number, we
shall briefly introduce our numerical model and then present and discuss 
our finding.

\red{We stress that the aim of our study is to draw a parallel between 
the dissipation anomaly in classical fluids which arises from
viscous effects and the dissipation anomaly in quantum turbulence which, 
as we shall see, arises at decreasing temperature from the
mutual friction \cite{friction} between the vortex lines and the normal fluid. 
Hence, in our numerical simulations of quantum turbulence,
the temperature must be high enough that the energy is indeed dissipated
by mutual friction at lengthscales larger than the vortex core radius, $a_0$,
and not by phonon emission (in $^4$He) or excitation of Carol-Matricon states 
(in $^3$He). These two effects would occur if
the energy cascade continued until the smallest scales of the flow 
($\approx a_0$). This is why our model is not the Gross-Pitaevskii
equation which has been found to describe the dissipation of incompressible
kinetic energy into phonons at zero temperature \cite{Nore1997}.} 

\section{Model}

\subsection{Superfluid Reynolds Number}
The first step is to identify the superfluid Reynolds number (a measure
of the intensity of the turbulence) by making careful analogy with
classical fluid dynamics. Classical fluids obey the 
Navier-Stokes equation. If $\nu$ is the kinematic viscosity of the fluid,
and $U$ and $L$ are characteristic speed and
length scales of the flow respectively, 
the dimensionless Navier-Stokes equation, written
in terms of the vorticity $\bom=\nabla \times \boldv$,
where $\boldv$ is velocity, has the form

\begin{equation}
\frac{\partial \bom}{\partial t}=\nabla \times (\boldv \times \bom) 
+\frac{1}{Re} \nabla^2 \bom,
\label{eq:NS}
\end{equation}

\noindent
where $Re=U L/\nu$ is the Reynolds number. The two terms at the 
right hand side of Eq.~\refeq{eq:NS} describe respectively
inertia and viscous dissipation.
Turbulence arises when $Re \gg 1$, \textit{i.e.} \blue{when inertia
is much larger than dissipation}.
In superfluid helium,
vorticity is not a continuous field but is concentrated in thin 
vortex lines of fixed atomic thickness, $a_0$, and fixed circulation,
$\kappa=h/m$,
where $h$ is Planck's constant and $m$ the mass of the relevant boson 
(an atom for bosonic $^4$He, a Cooper pair for fermionic $^3$He). 
The Hall-Vinen-Bekharevich-Khalatnikov (HVBK) equations 
\cite{HillsRoberts} provide a convenient coarse-grained, continuum model 
of finite temperature superfluid hydrodynamics
of fluid parcels threaded by a large number of vortex lines.
When the HVBK equations are applied to fully developed turbulence, 
vortex-tension effects are \blue{negligible} 
(being proportional to $1/Re_{\kappa}=\kappa/(U L) \ll 1$)
and the governing dimensionless 
equation for the superfluid vorticity $\bom_s$ is

\begin{equation}
\frac{\partial \bom_s}{\partial t}=(1-\alpha') \nabla \times 
(\boldv_s \times \bom_s) + \alpha \nabla \times [\bomhat_s \times
(\bom_s \times \boldv_s)],
\label{eq:HVBK}
\end{equation}

\noindent
where $\alpha$ and $\alpha'$ are known temperature-dependent
friction coefficients arising from the interaction of vortex lines 
with the normal fluid (which, for the sake of simplicity, we assume
to be at rest).
Following Finne {\it et al.} \cite{Finne} and the classical definition of
Reynolds number, we identify the superfluid Reynolds
number $Re_s$ as the ratio of inertial forces (the first term at the right hand
side of Eq.~\refeq{eq:HVBK}) to dissipative forces (the second term),
obtaining

\begin{equation}
Re_s=\frac{(1-\alpha')}{\alpha}.
\label{eq:Res}
\end{equation}

\noindent
Note that $Re_s$ does not depend on extrinsic parameters ($U$ and $L$)
but only on temperature-dependent fluid properties ($\alpha$ and $\alpha'$),
unlike the classical Reynolds number. 
We stress that experiments and numerical simulations \cite{Finne}
confirm that the transition
to turbulence can indeed be predicted using Eq.~(\ref{eq:Res}).

\subsection{Numerical Model}
The second step consists of numerical simulations of superfluid turbulence 
in which we compute the dissipation rate of turbulent kinetic energy,
$\epsilon$, as a function of $Re_s$. We employ the
well-established vortex filament method (VFM) \cite{Schwarz,HanninenBaggaley} 
which models superfluid hydrodynamics at length 
scales smaller than the average inter-vortex distance, $\ell$, but 
much larger than the vortex core radius, $a_0$. 
Unlike the HVBK framework, the VFM
still describes the discrete nature of superfluid vorticity.
Vortex lines are \blue{described} as one-dimensional filaments, 
$\bolds(\xi,t)$, 
$\xi$ being the arc length and $t$ time, which move according to the 
balance of Magnus and friction forces.
The velocity of a vortex line at point $\bolds(\xi,t)$ is 

\begin{equation}
\frac{\partial\bolds}{\partial t}=\boldv_s-\alpha \bolds' \times \boldv_s 
+\alpha' \bolds' \times (\bolds' \times \boldv_s),
\label{eq:Schwarz}
\end{equation}

\noindent
where

\begin{equation}
\boldv_s(\bolds(\xi,t),t)=\frac{\kappa}{4 \pi} \oint_{\mathcal{T}}
\frac{\bolds'(\xi_1,t)\times \left ( \bolds(\xi,t) - \bolds(\xi_1,t) \right )}{\vert \bolds(\xi,t) - \bolds(\xi_1,t) \vert^3} d\xi_1,
\label{eq:BS}
\end{equation}

\noindent
the line integral (desingularized \blue{as in} \cite{schwarz-1985})
extending over the entire vortex configuration $\mathcal{T}$, and
$\bolds'=\partial \bolds/\partial \xi$ being the unit tangent at $\bolds$. 
The local curvature (the inverse of the local radius of curvature) is
defined as $\zeta=\vert \bolds''\vert$, where 
$\bolds''=\partial^2 \bolds/\partial \xi^2$.
In the VFM, 
each filament is
discretized into a variable number of oriented line elements held at a
distance $\Delta \xi \in [\delta/2,\delta ]$; \blue{here}
we choose $\delta=0.02\, \rm{cm}$, 
and check results by halving $\delta$.
Each simulation is performed at \blue{temperature} $T$ 
in a periodic cube of size $D=1~\rm cm$. 
The time integration is a Runge-Kutta 4th order scheme, 
$\Delta t=5 \times 10^{-3}~\rm s$ being the typical time step.
Reference~\cite{HanninenBaggaley} gives more details, including 
vortex reconnections performed algorithmically 
when two filaments collide.

A fully realistic model of turbulent $^4$He would need to
couple Eq.~\refeq{eq:Schwarz} to a Navier-Stokes equation for the turbulent
normal fluid velocity $\boldv_n$, suitably modified to include 
the friction arising from the 
relative motion of vortex lines and $\boldv_n$. Due 
to \blue{computational costs, such}
coupled dynamics of vortex lines and normal fluid has been attempted  
only for particular vortex 
configurations \cite{Kivotides2000,Galantucci2020,Galantucci2022} 
or turbulent transients \cite{Kivotides2007,Morris2008}:
fully developed, statistically-steady two-fluid turbulence 
has not been achieved yet.
Therefore, we limit this investigation 
to the following idealised but computationally simpler form of 
superfluid turbulence: a tangle of vortex lines whose dissipative motion
with respect to a quiescent normal fluid is modelled by 
Eq.~\refeq{eq:Schwarz} at the mesoscale level. 
This simpler form still allows us to make progress into a
problem never addressed before. It is worth stressing that although 
idealised for $^4$He, our model is realistic for $^3$He-B, whose normal 
component is so viscous that it is usually assumed to be at rest with 
respect to the container (see Appendix \ref{appx:A}).

\section{Results}
\begin{figure}[!htbp]
\centering
\begin{minipage}[c]{0.40\columnwidth}
\includegraphics[angle=0,width=1\textwidth]{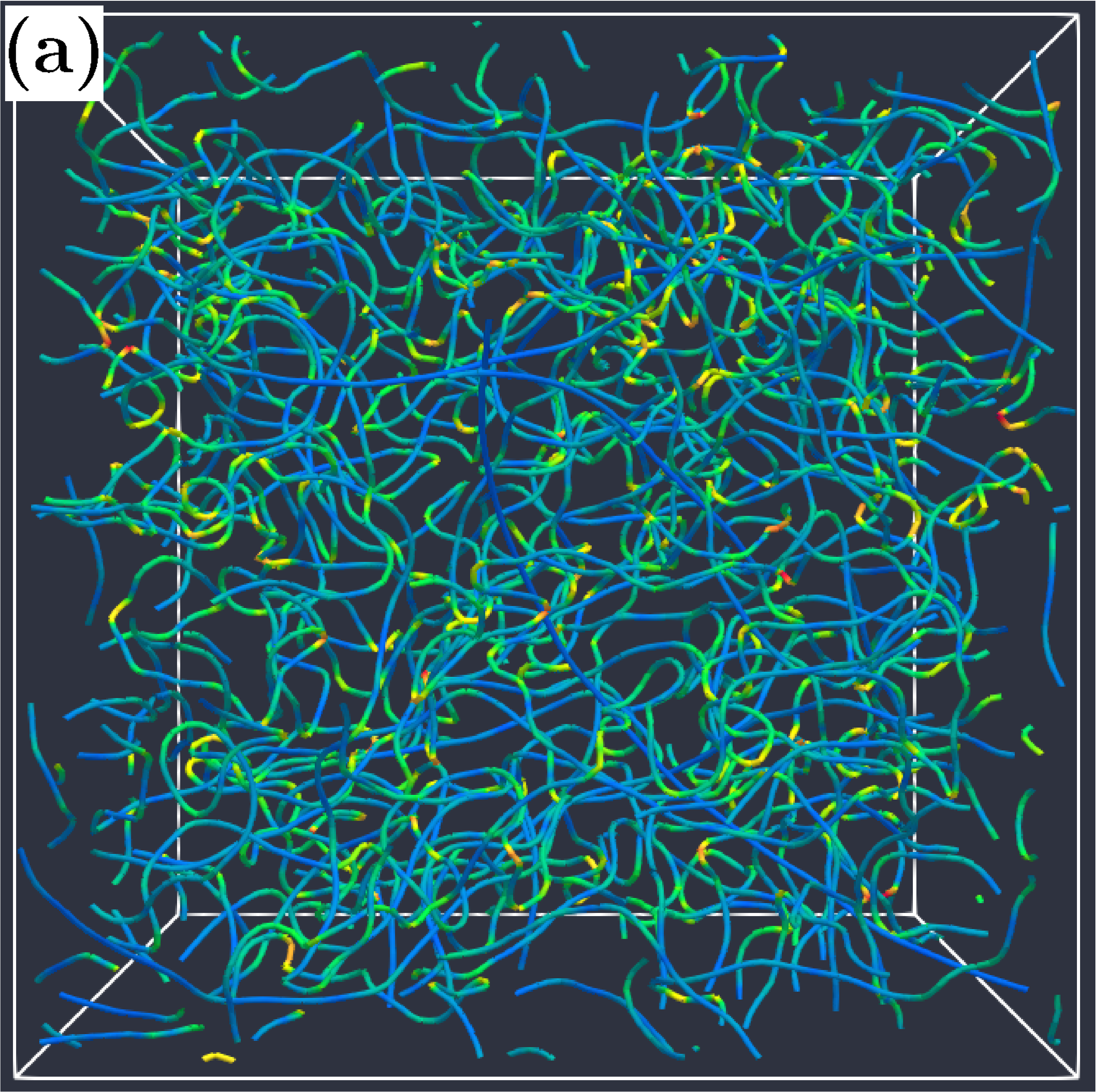}
\end{minipage}
\hspace{0.10\columnwidth}
\begin{minipage}[c]{0.40\columnwidth}
\includegraphics[angle=0,width=1\textwidth]{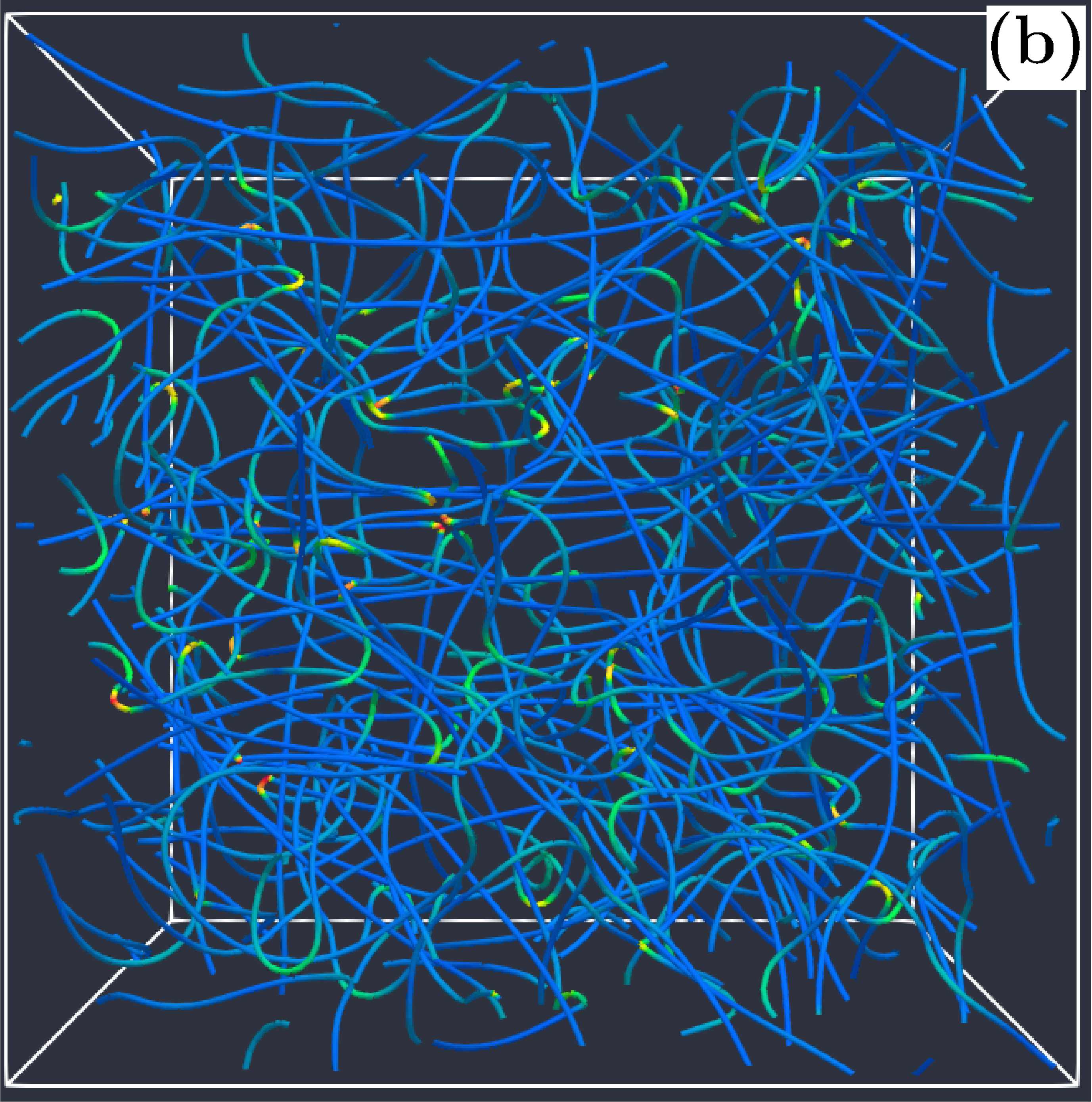}
\end{minipage}\\[3mm]
\begin{minipage}[c]{0.30\columnwidth}
\end{minipage}
\hspace{-0.03\columnwidth}
\begin{minipage}[c]{0.30\columnwidth}
\includegraphics[angle=0,width=1\textwidth]{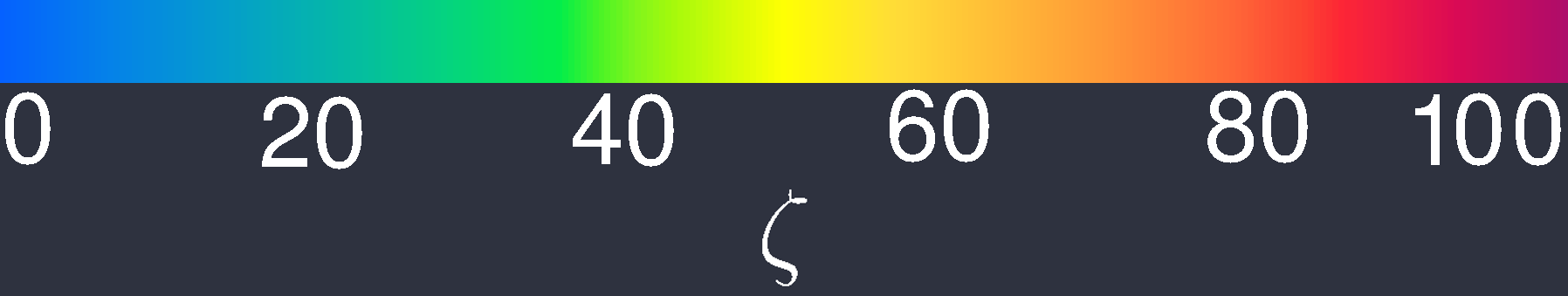}
\end{minipage}
\caption{
Snapshot of vortex tangles in the saturated steady-state
regime at $\overline{L}\sim 120 \rm cm^{-2}$, for: (a) $Re_s=29$ and ${\dot L}_{inj}=3.35 \rm cm^{-2}  \rm s^{-1}$
(corresponding \blue{to pink open diamond symbols} in Fig.~\ref{fig:2} (a));
(b) $Re_s=1.25$ and ${\dot L}_{inj}=22.50 \rm cm^{-2}  
\rm s^{-1}$
(corresponding to \blue{violet filled diamond symbols} 
in Fig.~\ref{fig:2} (a)). 
Vortex lines are colour-coded according to the local curvature $\zeta$ (in $\rm cm^{-1}$,
legend at the bottom); note the larger values of $\zeta$
at the larger $Re_s$.
}
\label{fig:1}
\end{figure}

To establish a turbulent flow, 
we start with a vortex ring of radius $R=D/2$ at
the centre of the box, and inject 
randomly oriented vortex rings of the same radius $R$ at
random positions and at a prescribed rate ${\dot L}_{inj}$
(a similar procedure was used in the experiment of 
Walmsley and Golov \cite{Walmsley2008}, although
their injection was not isotropic).
The injected vortex rings interact with other vortex lines, 
reconnect, and a turbulent vortex tangle is formed,
see Fig.~\ref{fig:1}~(a,b).

\begin{figure}[!htbp]
\centering
\begin{minipage}[c]{0.40\columnwidth}
\includegraphics[angle=0,width=0.95\textwidth]{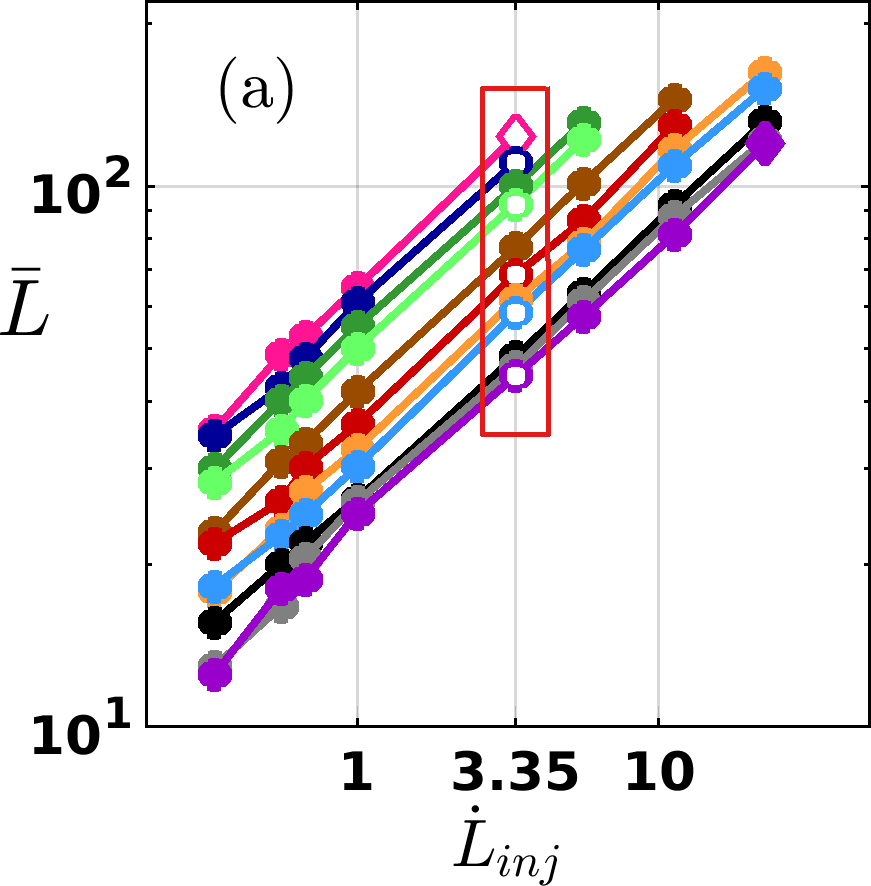}
\end{minipage}
\hspace{0.10\columnwidth}
\begin{minipage}[c]{0.40\columnwidth}
\includegraphics[angle=0,width=1\textwidth]{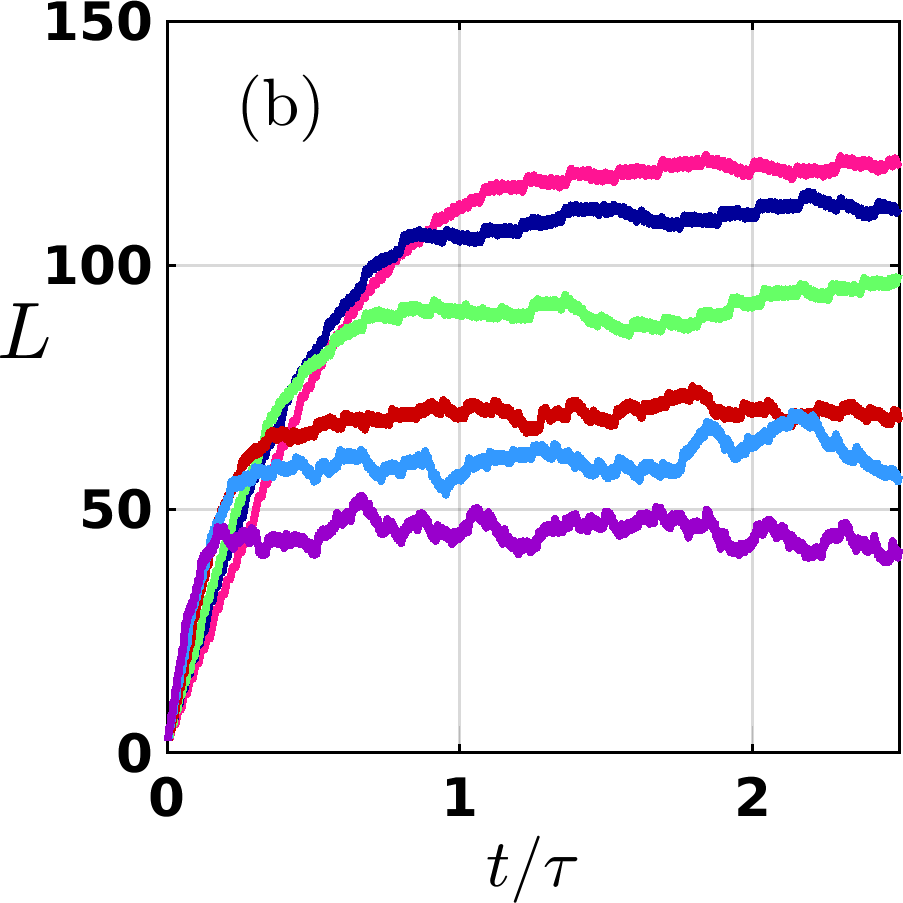}
\end{minipage}\\
\caption{
(a): Vortex line density at saturation, $\overline{L}$ (in $\rm cm^{-2}$),
\textit{vs} injection rate $\dot{L}_{inj}$ (in $\rm cm^{-2}  \rm s^{-1}$ )
for $Re_s=$1.25 (violet), 1.34 (grey), 
1.50 (black), 2.06 (cyan), 2.51 (yellow), 3.30 (red), 4.96 (brown), 
9.84 (light green), 13.09 (dark green), 19.66 (blue), 29.00 (pink).
(b): Time evolution of the vortex line density $L$ (in $\rm cm^{-2}$)
\textit{vs} time $t/\tau$, 
at $\dot{L}_{inj}=\, 3.35 \; \rm cm^{-2}  \rm s^{-1}$, for different values of $Re_s$ (colors
as in (a)), corresponding to simulations indicated with open symbols in (a).
}
\label{fig:2}
\end{figure}

Without continual injection, the tangle would decay due to the 
friction suffered by the vortex lines as they move in
the quiescent normal fluid background.  The statistically-steady state of turbulence 
which is achieved after an initial transient $T_{eq}$ is
independent of the initial condition (injection and dissipation
balancing each other). In this state, the vortex line density $L$ 
(defined as the vortex line length per unit volume) fluctuates 
around a constant saturation value $\overline{L}$, as illustrated
in Fig.~\ref{fig:2}~(b).
The diameter of the injected rings is equal to the box size and hence 
energy is mainly supplied to the turbulence at scales
larger than $\ell$ \cite{Leonard1985,Yurkina2021} (see also Appendix \ref{appx:C}). 
As $\boldv_n$ is kept static, mutual friction dissipates 
energy at all lengthscales, implying that the turbulence \blue{thus
generated} is not quasi-classical 
(where by the latter we intend a quantum turbulent flow where at 
large scales the two 
fluid components are coupled by mutual friction and
hence undergo a coupled energy cascade with little dissipation 
until a lengthscale~$\sim~\ell$ is reached
\cite{BLR2014,walmsley-etal-2014b,SkrbekSreeni}).   
However, at scales larger than $\ell$, we still do observe the 
emergence of an inertial-range 
energy cascade and the subsequent Kolmogorov
energy spectrum $\hat{E}(k) \sim k^{-5/3}$ (see Fig.~\ref{fig:3_bis} (a) and Appendix~\ref{appx:C}), 
as the largest  eddy turnover time $\tau_D=(D^2/\epsilon)^{1/3}$ 
is never significantly larger than the mutual friction 
dissipative time-scale 
$\tau_{mf}=1/(\alpha \kappa \overline{L})$. 
\blue{Our computational box is not large enough that $\tau_D \gg \tau_{mf}$, 
which would make friction dominant at large scales creating a crossover 
to a $k^{-3}$ scaling 
\cite{vinen-2005,lvov-nazarenko-volovik-2004,SkrbekSreeni,Skrbek2021}.
%
}

When in our simulations the mentioned Kolmogorov energy cascade 
reaches scales~$\sim~\ell$,
the energy transfer towards smaller scales $k \gtrsim k_\ell = 2\pi/\ell$
creates Kelvin waves (KWs) of shorter and shorter wavelengths
on individual vortex lines.  In our temperature range 
($T\ge 1.3$K), this KW cascade \cite{KozikSvistunov,LvovNazarenko,Krstulovic} 
is limited by the friction with the 
normal fluid \cite{vinen-2005,walmsley-etal-2014b,vinen-niemela-2002}.

In order to assess the effect of turbulent intensity on 
the rate of dissipation 
of kinetic energy, $\epsilon$, we choose  11
\blue{temperature} values in the range
$1.3~{\rm K} \le T \le 2.16~{\rm K}$, and 8 injection rates
$\dot{L}_{inj}=(dL/dt)_{inj}$ in the range
$ \displaystyle 0.34~\rm{cm}^{-2} \rm{s}^{-1} \le \dot{L}_{inj} \le 22.50~
\rm{cm}^{-2} \rm{s}^{-1}$.
The saturation value, $\overline{L}$, increases with $\dot{L}_{inj}$ and
decreases with $T$.  At saturation, on average 
$\dot{L}_{tot}=\dot{L}_{inj} + \dot{L}_{decay}=0$, 
leading to $\overline{L} = \sqrt{2\pi\dot{L}_{inj}/(\kappa \chi_2)}$ \cite{vinen-1957c}, 
%
%
as confirmed by our numerical simulations, see Fig.~\ref{fig:2} (a). The values of $\chi_2$ extrapolated from our data are 
consistent with recent studies \cite{gao-etal-2018}.

To extract values of the energy dissipation rate 
$\epsilon$ as a function of $Re_s$ we select the numerical simulations 
corresponding to a constant
value of $\dot{L}_{inj}$ 
(we choose $\dot{L}_{inj}=\, 3.35 \; \rm cm^{-2}  \rm s^{-1}$), 
implying that the only varying physical parameter among the distinct 
simulations is $T$ (or, equivalently, $Re_s$).
The selected simulations are
enclosed in the red rectangle in Fig.~\ref{fig:2} (a)) and 
the temporal evolution of the vortex line density $L$ for this set of simulations at constant $\dot{L}_{inj}$ 
is illustrated in Fig.~\ref{fig:2} (b). For each $Re_s$, after the transient $T_{eq}$, 
we calculate the dissipation rate $\epsilon$ every time interval $\tau=2\pi/(\kappa \overline{L})$ 
via the following integral,
\begin{equation}
\displaystyle
\epsilon(t_i)=\frac{1}{\rho_s D^3} \oint_{_{\mathcal{T}(t_i)}}\!\!\!\! \bold{f}_{ns} (\xi,t_i)\! \cdot\! \dot{\bf{s}} (\xi,t_i) \text{d}\xi \; \; ,
\label{eq:epsilon_REV}
\end{equation}
where $- \bold{f}_{ns}=-\gamma_0 \, \dot{\bf{s}} - \gamma_0' \, \bolds' \times \dot{\bf{s}}$ is the mutual friction force per unit length 
which the normal fluid exerts on a superfluid vortex line 
element \cite{BDV1983} 
($\gamma_0$ and $\gamma_0'$ being a reformulation of mutual friction coefficients $\alpha$ and $\alpha'$),
and $\mathcal{T}(t_i)$ is the configuration of vortex tangle at time $t_i=T_{eq}+i\cdot \tau$
(with $i=1, \dots , 10$).

\begin{figure}[!htbp]
\centering
\includegraphics[angle=0,width=0.70\columnwidth]{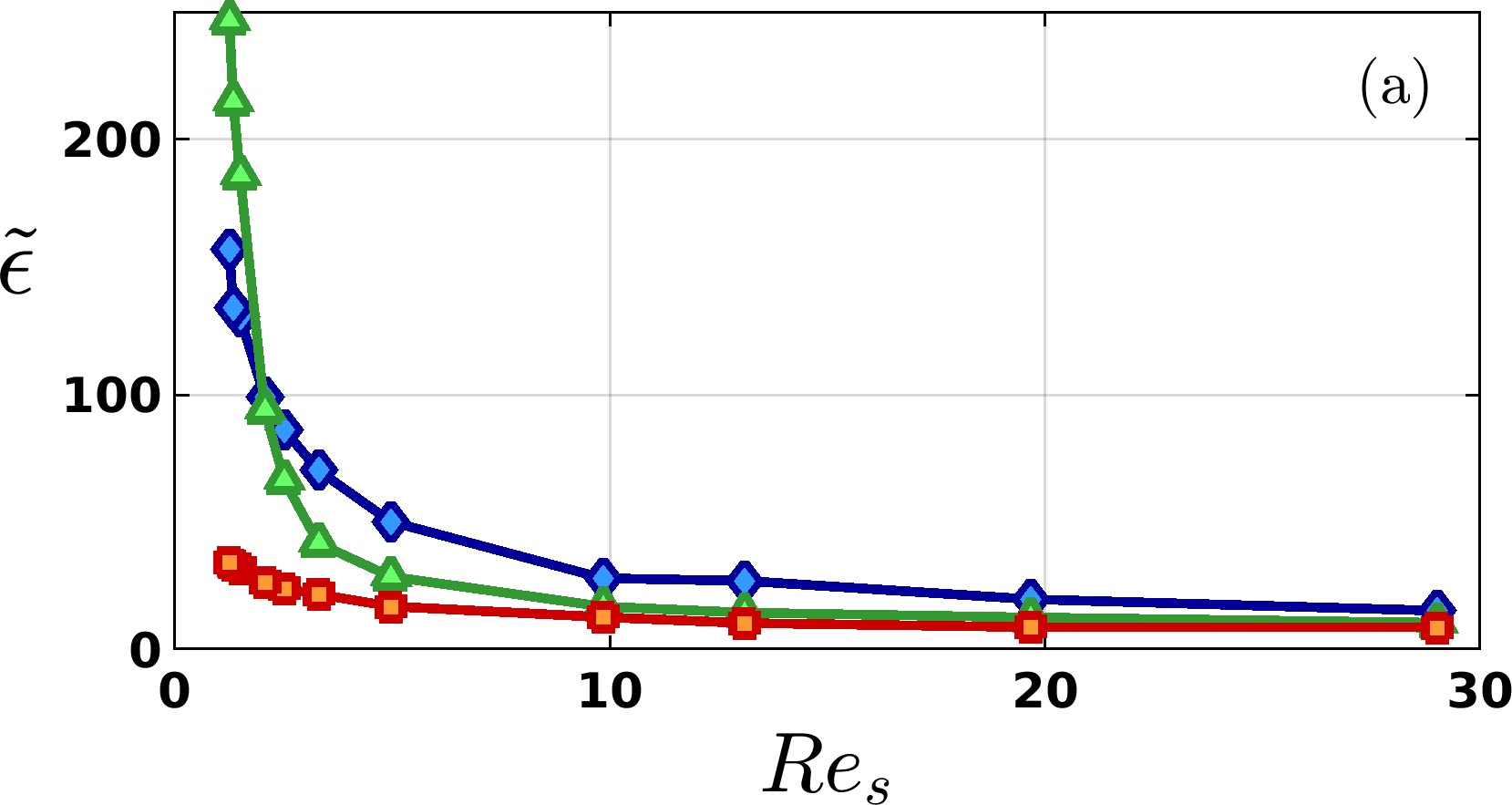}\\[1mm]
\includegraphics[angle=0,width=0.75\columnwidth]{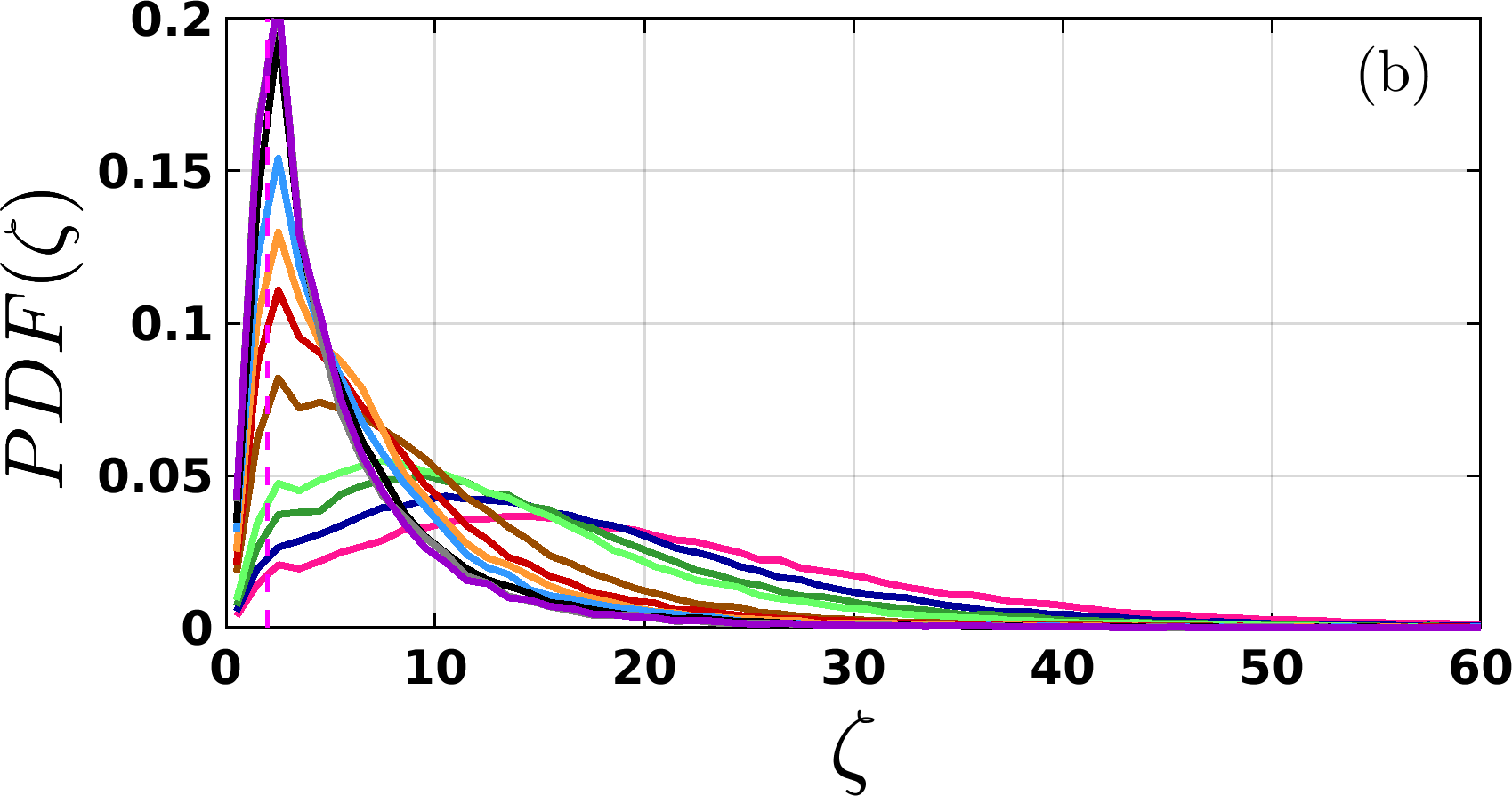}
\caption{
(a): Normalised energy dissipation rate,
${\widetilde\epsilon}$, \textit{vs} superfluid Reynolds number, $Re_s$, 
at $\dot{L}_{inj}=3.35 \rm cm^{-2}  \rm s^{-1}$. 
Red curve: injected vortex ring radius $R=D/2$; Green curve: $R=D/8$;
Blue curve: $R\gtrsim \bar{\ell}/2$.
(b): probability density function  
of the curvature ${\rm PDF}(\zeta)$ (in $\rm cm$) \textit{vs} curvature $\zeta$ (in cm$^{-1}$) at increasing 
$Re_s$, for $\dot{L}_{inj}=3.35 \rm cm^{-2}  \rm s^{-1}$ and $R=D/2$. Colors as in Fig.~\ref{fig:2}. The vertical dashed
magenta line marks the curvature of the injected rings $\zeta_0=2/D$.}
\label{fig:3}
\end{figure}

At the same times $t_i$, we evaluate the root-mean-square velocity fluctuation $U(t_i)$, where 
$3 U^2(t_i) /2 = E(t_i)$ is the turbulent kinetic energy per unit mass, and the turbulent integral scale
$\displaystyle I(t_i)=\frac{\pi}{2 U^2}\int_0^\infty \frac{\hat{E}(k,t_i)}{k}\;\text{d}k$, where 
$\hat{E}(k,t_i)$ is the one-dimensional energy spectra so that $E(t_i)=\int_0^\infty \hat{E}(k,t_i)\;\text{d}k$.
Finally, we average over time $\epsilon$, $U$ and $I$ and compute the normalised energy dissipation rate 
as $\widetilde{\epsilon}=\langle \epsilon \rangle \langle I \rangle/\langle U \rangle^3$ ($\langle \cdot \rangle$
indicating time-averaged quantities), in order to mimic the normalisation performed in classical turbulence. This 
procedure is repeated for each $Re_s$ and the curve $\widetilde{\epsilon}(Re_s)$ is plotted in Fig.~\ref{fig:3} (a), red curve.

Fig.~\ref{fig:3} (a) shows our main result:
the normalised energy dissipation rate
$\widetilde{\epsilon}$ decreases and then flattens out as
the superfluid Reynolds number, $Re_s$, is increased:
the similarity with the classical dissipation anomaly 
\cite{Sreeni1984,KanedaIshihara2003} is striking.

\subsection{Interpretation of the results}
\blue{
We examine} the geometry of the vortex tangle.
Fig.~\ref{fig:3} (b) shows the  
probability density function (PDF) of the curvature 
$\zeta$ along the vortex lines as a function of $Re_s$.
Clearly, increasing $Re_s$ shifts this distribution towards 
higher $\zeta$, hence towards smaller length scales $1/\zeta$. 
The small-scale (large $\zeta$) vortex structures
generated at lower temperatures 
survive because of the reduced friction dissipation
\red{and, as a consequence, the probability of observing structures at scales
smaller than $\ell$ increases as $Re_s$ increases (see Appendix \ref{appx:PDF_larger_ell}).}
The energy cascade towards small scales
can be described as a shift of ${\rm PDF}(\zeta)$ towards high curvatures, 
starting from the injected value 
$\zeta_0=1/R$ (see Appendix \ref{appx:E}).

\begin{figure}[!htbp]
\begin{minipage}[c]{0.40\columnwidth}
\includegraphics[angle=0,width=1\textwidth]{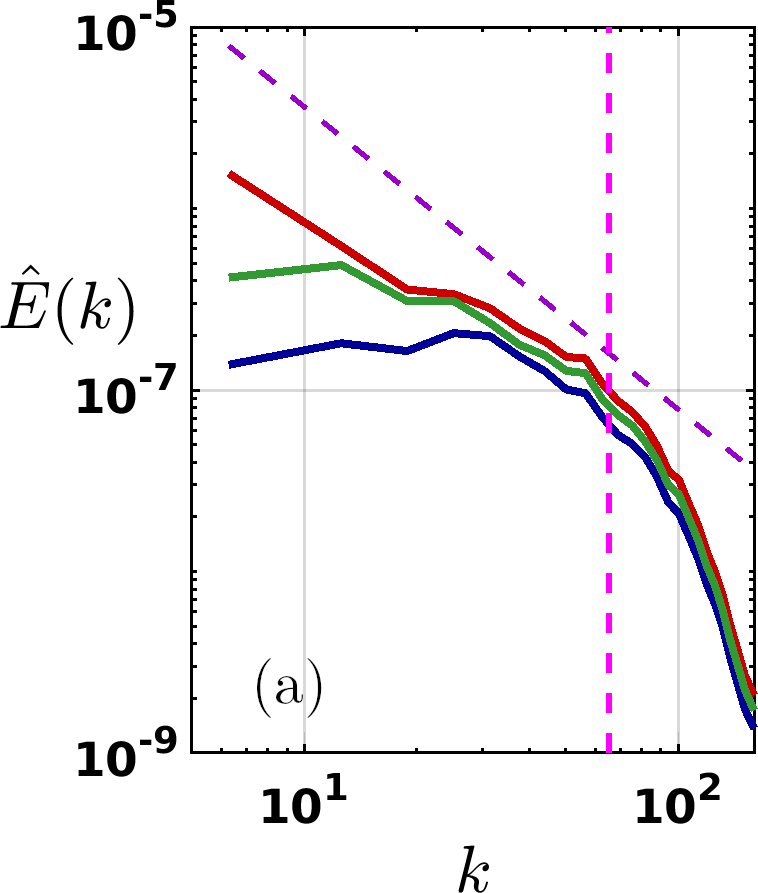}
\end{minipage}
\hspace{0.10\columnwidth}
\begin{minipage}[c]{0.40\columnwidth}
\includegraphics[angle=0,width=1\textwidth]{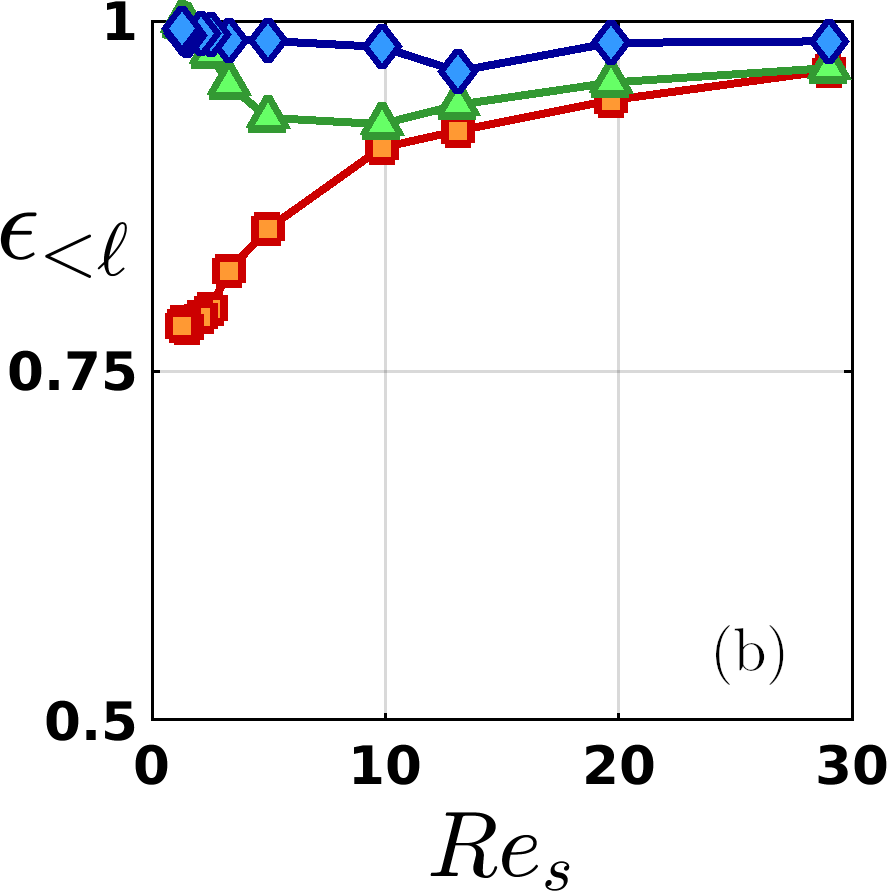}
\end{minipage}\\
\caption{(a): time averaged superfluid kinetic energy 
spectra $\hat{E}(k)$ (arbitrary units) as a function
of wavenumber $k$ (cm$^{-1}$) at saturation 
for $\dot{L}_{inj}=3.35 \rm cm^{-2}\rm s^{-1}$ and $Re_s=29$ ($\hat{E}(k)$ for lower $Re_s$ in Appendix~\ref{appx:C}). 
Vertical pink and oblique violet dashed lines indicate 
$k_\ell$ and the $k^{-5/3}$ scaling, respectively.
(b): fraction $\epsilon_{<\ell}$ of total dissipation arising from lengthscales smaller than 
$\ell$. Colors, as in Fig.~\ref{fig:3} (a) refer to
different radii of the injected vortex rings.
}
\label{fig:3_bis}
\end{figure}

As in classical turbulence, smaller \blue{friction}
(decreasing values of $\alpha$) \blue{leads} to
the excitation of smaller scale motions. 
The flattening of the $\widetilde \epsilon$
curve can be understood using the following simple argument \violet{which
is only strictly valid for Vinen turbulence, but 
is likely to be applicable, at least qualitatively,
to other quantum turbulent regimes, as the dissipation stems from the small-scale dynamics
(see Fig.\ref{fig:3_bis} (b) and subsequent discussion), independently of the large scale flow features.}
The kinetic energy per unit mass
$f(t)$ of a vortex ring of radius $R$ at time $t$ is 
$f(t) \sim \kappa^2 R(t)/\bar{\ell}^3$ (where $\bar{\ell}$ is the average 
inter-vortex spacing at saturation).
The dissipation rate $\epsilon$ is hence given by

\begin{equation} 
\epsilon=-df/dt=- \kappa^2\dot{R}/\bar{\ell}^3 = \alpha \kappa^3 \zeta/\bar{\ell}^3 \simeq \alpha \kappa^3 \zeta^4 \, \, , \label{eq:epsilon_ring}
\end{equation}

\noindent
where \violet{we have employed}
the well-known shrinking rate of a vortex ring 
in a quiescent normal fluid, $\dot{R}\sim -\alpha\kappa/R$ (cf. Eq.~(\ref{eq:Schwarz})), \violet{and
the relation $\langle \zeta^2 \rangle \propto \bar{\ell}^{-2}$ \cite{Schwarz}.}
As $Re_s$ increases, the decreasing value of $\alpha$ 
is thus compensated by larger curvatures $\zeta$
on the vortex lines,  
flattening $\widetilde{\epsilon}$ as shown in Fig.~\ref{fig:3} (a) (red curve).
The presence of larger values of $\zeta$ (smaller structures) 
along the vortex lines
as $Re_s$ increases is clearly visible in Fig.~\ref{fig:1}~(a,b).
This behaviour is analogous to the scenario observed in classical 
turbulence where the dissipation
rate $\epsilon_{class}~=~(\nu/2)(\partial v_i/\partial x_j + \partial v_j/\partial x_i)^2$ tends to a finite constant as 
decreasing viscosity is balanced by increasing velocity gradients. 
Our results hence
show that the curvature of vortices in quantum turbulence play the same role of enstrophy in classical turbulence,
as, in terms of dissipative effects, small-scale one-dimensional structures on superfluid vortices correspond to
the classical dissipative eddies.


To evaluate the importance of the smallest scales ($k>k_\ell$) 
in dissipating the superfluid kinetic energy,
we calculate the fraction $\epsilon_{<\ell}$ of total dissipation arising from the motion of vortexline elements 
with curvature $\zeta>1/\ell$. The result is illustrated in Fig.~\ref{fig:3_bis} (b) (red curve) where we observe
that $\epsilon_{<\ell}$ is larger than $0.75$ for all $Re_s$. 
\dgreen{The value of $\epsilon_{<\ell}$ close to $1$ for the largest $Re_s$ and its 
slight decrease for decreasing $Re_s$ stems from the fact that $\tau_{mf}\rightarrow\tau_D$ as $Re_s\rightarrow 0$, consistently
with previuos theoretical predictions \cite{vinen-2005}.}

This predominant role played by the smallest scales in the dissipation
implies that the superfluid analog to the classical dissipation anomaly 
\red{does not depend on}
the mechanism transferring 
the energy to such small scales. To show this independence from 
the largest scales, we repeat our numerical
experiment injecting smaller rings of radii $R=D/8$ 
and $R\gtrsim \bar{\ell}/2$.
These injection protocols produce 
Vinen-like energy spectra which peak at intermediate scales
\cite{Baggaley2012}, as shown 
\red{in Fig.~\ref{fig:3_bis} (a, green and blue curves).}
Despite this non-classical aspect
at large scales, the dissipation anomaly is still clearly 
evident, \red{ see Fig.~\ref{fig:3} (a, green and blue curves).
This result does not depend on the normalisation, as shown in Appendix~\ref{appx:D}.}

\subsection{Numerical resolution of the small length scales}
As increasing $Re_s$ excites smaller length scales along the vortex lines, 
it is natural to ask whether our numerical discretization correctly 
resolves these small scales.  
To assess our numerical resolution we have repeated all the
simulations replacing $\delta$ with $\delta/2$ and in the
calculation of $\widetilde{\epsilon}$ we have
rejected the results of simulations which do not satisfy strict 
criteria regarding the saturation value $\overline{L}$ and 
curvature $\zeta$ (see Appendix \ref{appx:B}). 
In practice, our strict criterion limits us to temperatures above 
$T \approx 1.3~\rm K$, above the appearance of scaling behaviour
for the KW cascade \cite{KozikSvistunov,LvovNazarenko,Krstulovic}.
\red{Therefore, our model does not suffer the
numerical dissipation at the small
length scales which occurs in the VFM if the temperature is set to zero
\cite{Araki2002}.}


\section{Conclusions}
\red{Our numerical investigation shows that to understand the
small-scale dynamics of superfluid turbulence one has to consider
the full distribution of the curvature along the vortex lines,
not simply the average value.}
We have shown that, superfluid turbulence displays
the same dissipation anomaly which is observed in 
classical turbulence: 
the effect of increasing Reynolds number is the creation of
smaller length scales. 
This result 
concerning the smallest length scales of turbulence adds insight
into the remarkable analogies between classical turbulence and
superfluid turbulence already noticed at the largest length scales
\cite{BLR2014,SkrbekSreeni}. It is a striking result, because
it \blue{deals with} lengthscales smaller than the average inter-vortex
distance, where classical and quantum turbulence have always been 
believed to differ \cite{Skrbek2021}.
The role of the quantisation of circulation is thus to \textit{constrain} 
these dissipative
structures to live on vortex lines rather than in the bulk of the flow.

Our results illustrate the nature and the dynamical origin of the 
recent observation
of a dissipation anomaly obtained by forcing KWs
in superfluid $^3$He-B \cite{makinen-etal-2022}, 
contributing to the lively debate regarding in which turbulent systems
dissipative anomaly manifests itself \cite{john-etal-2021}.

\begin{acknowledgments}
{\it Acknowledgments.---}
We thank Giorgio Krustulovic and Ladislav Skrbek for useful discussions, and
acknowledge the support of EPSRC grant EP/R005192/1.
LG acknowledges the support of Istituto Nazionale di Alta Matematica (INdAM).
\end{acknowledgments}

\appendix


\section{Application to $^3$He}\label{appx:A}

In the numerical simulations with the VFM
we use values of parameters relative to $^4$He: 
$\kappa=h/m=9.97 \times 10^{-7}~\rm m^2/s$ (where $h$ is Planck's
constant and $m$ the mass of one atom of $^4$He),
$a_0 \approx 10^{-10}~\rm m$, $\alpha$ and $\alpha'$ from Ref.~\cite{DB}.
However our main conclusions are also relevant to
$^3$He-B for the following reasons.
\bigskip

\noindent (i):
In $^3$He the relevant boson is a Cooper pair consisting of two $^3$He atoms
(each having mass equal to $3/4$ of $m$), 
therefore the quantum of circulation is $2/3$ of the value in $^4$He. 
This difference implies
a small rescaling of the characteristic velocity, hence of time, for example
when judging the duration of numerical simulations, such as the simulations reported in 
Fig.~\ref{fig:2} (b).
\bigskip

\noindent (ii):
The different values of the friction coefficients imply a simple
rescaling of $T$, hence of $Re_s$ in Figs.~\ref{fig:3} (a) and \ref{fig:3_bis} (b).
\bigskip

\noindent (iii):
The mesoscopic length scales described by the VFM are much larger 
than the vortex core radius 
in both $^4$He and $^3$He ($a_0 \approx 10^{-6}\rm cm$); 

\section{Energy spectra}\label{appx:C}

\begin{figure}[ht]
\centering
\begin{minipage}[c]{0.35\columnwidth}
\includegraphics[angle=0,width=1\textwidth]{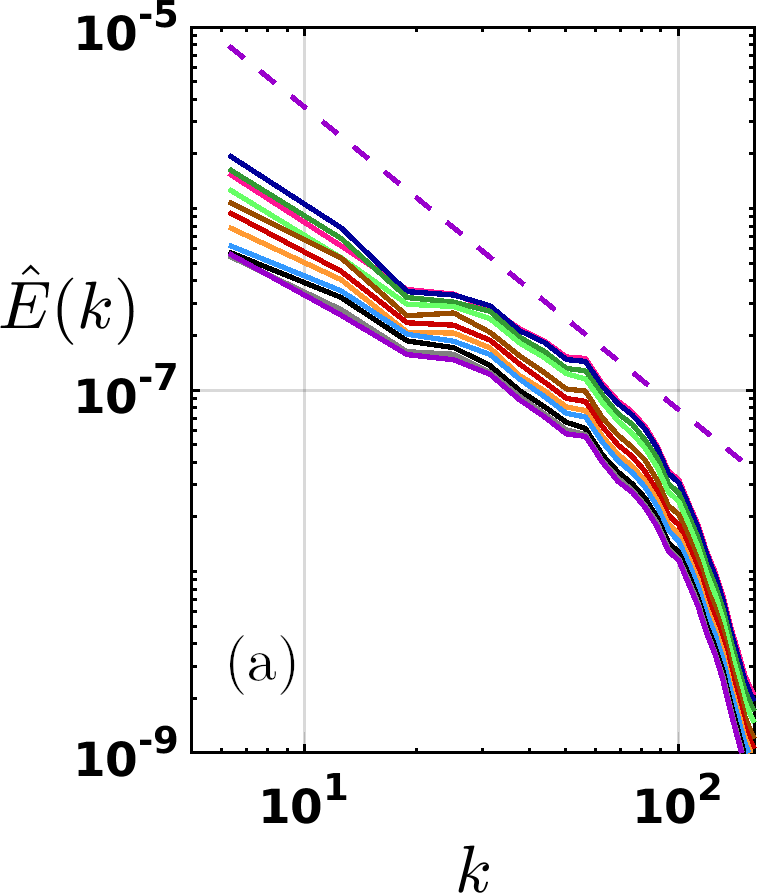}
\end{minipage}
\hspace{0.10\columnwidth}
\begin{minipage}[c]{0.35\columnwidth}
\includegraphics[angle=0,width=1\textwidth]{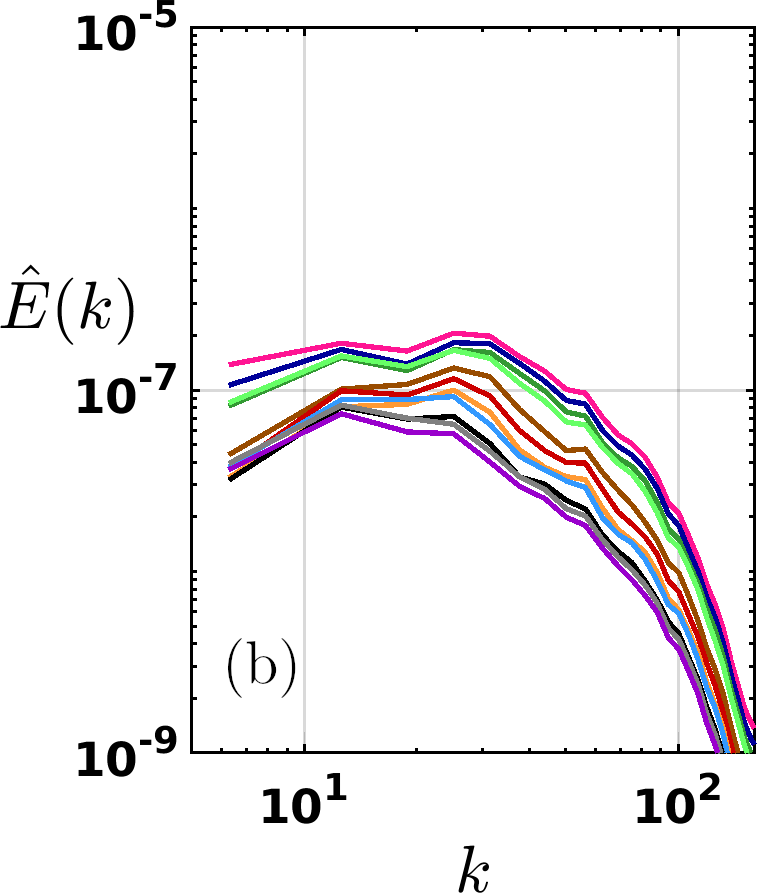}
\end{minipage}
\caption{Superfluid kinetic energy time-averaged spectra $\hat{E}(k)$ (arbitrary units) as a function
of wavenumber $k$ (cm$^{-1}$) for  $\dot{L}_{inj}=3.35 \rm cm^{-2}  \rm s^{-1}$. 
Different colors refer to distinct values of $Re_s$: 1.25 (violet), 1.34 (grey), 
1.50 (black), 2.06 (cyan), 2.51 (yellow), 3.30 (red), 4.96 (brown), 
9.84 (light green), 13.09 (dark green), 19.66 (blue), 29.00 (pink).
(a): radius of injected rings $R=D/2$; (b): radius of injected rings $R\gtrsim \bar{\ell}/2$.
The dashed violet curve in (a) indicates the Kolmogorov $k^{-5/3}$ energy spectrum.
}
\label{fig:S1}
\end{figure}

In this Appendix \ref{appx:C} we illustrate the behaviour
of the time-averaged energy spectra $\hat{E}(k)$ for different injected ring  
radii. In Fig.~\ref{fig:S1} (a) and (b) we report
the time-averaged spectra $\hat{E}(k)$ \textit{vs} wavenumber $k$ for injected ring radii $R=D/2$  and $R\gtrsim \bar{\ell}/2$, respectively.
The distinct curves reported in Fig.~\ref{fig:S1} (a) and (b) correspond to all
values of $Re_s$ employed in the numerical simulations (the color legend coincides
with the legend used in Fig.~\ref{fig:2}). The injection rate
$\dot{L}_{inj}=3.35 \rm cm^{-2}  \rm s^{-1}$ is fixed. 

In Fig.~\ref{fig:S1} (a) we observe that when the ring is injected at the largest scales of the flow, the energy spectrum $\hat{E}(k)$
is precisely peaked at scale $D$. In addition, at these large scales we can observe the emergence of a Kolmogorov $k^{-5/3}$
spectrum. As illustrated in the main manuscript, this Kolmogorov spectrum does \textit{not} imply 
that the quantum turbulence that we generate is quasi-classical 
(where by the latter we intend a quantum turbulent flow where at large scales the two 
fluid components are coupled by mutual friction and
hence undergo a coupled energy cascade with little dissipation until a lengthscale~$\sim~\ell$ is reached).
As the normal fluid is in fact kept quiescent, the mutual friction force acts at all lengthscales. However, 
given that the mutual friction characteristic time-scale is never sufficiently small when compared to the 
eddy turnover time, we still observe an the emergence of an inertial-range Kolmogorov spectrum at all $Re_s$. 
\cite{vinen-2005,lvov-nazarenko-volovik-2004,SkrbekSreeni,Skrbek2021}.

On the other hand, in Fig.~\ref{fig:S1} (b), as the energy is injected at scales comparable to the 
average inter-vortex spacing at saturation, we observe Vinen-like (often also called ultra-quantum) energy spectra,
peaked at intermediate lengthscales, for all values of $Re_s$.

\newpage

\section{Probability of observing scales smaller than $\ell$}\label{appx:PDF_larger_ell}

\red{
In this section of the Appendix, in Fig.~\ref{fig:R1} (top) we report the Probability Density Function of the 
curvature $PDF(\zeta)$ for selected values of $Re_s= 1.25, \, 2.5, \, 9.8, \, 29$,
indicating the corresponding value $\zeta_\ell=1/\ell$, which increases for increasing $Re_s$. This
figure is almost identical to Fig.~\ref{fig:3} (a), the only differences being the the indication of $\zeta_\ell$
and the selection of fewer values of $Re_s$ in order to ease the readability of the figure. 
On the basis of this data, we have calculated the 
probability $P(\zeta > \zeta_\ell)$ of observing structures
at length scales smaller than $\ell$ as a function of the superfluid Reynolds number $Re_s$, reporting
the results in Fig.~\ref{fig:R1} (bottom).
Fig.~\ref{fig:R1}
shows that structures at scales smaller than $\ell$ exist and that
the probability of observing such small structures increases 
as $Re_s$ increases.}

\begin{figure}[!htbp]
\centering
\includegraphics[angle=0,width=0.6\textwidth]{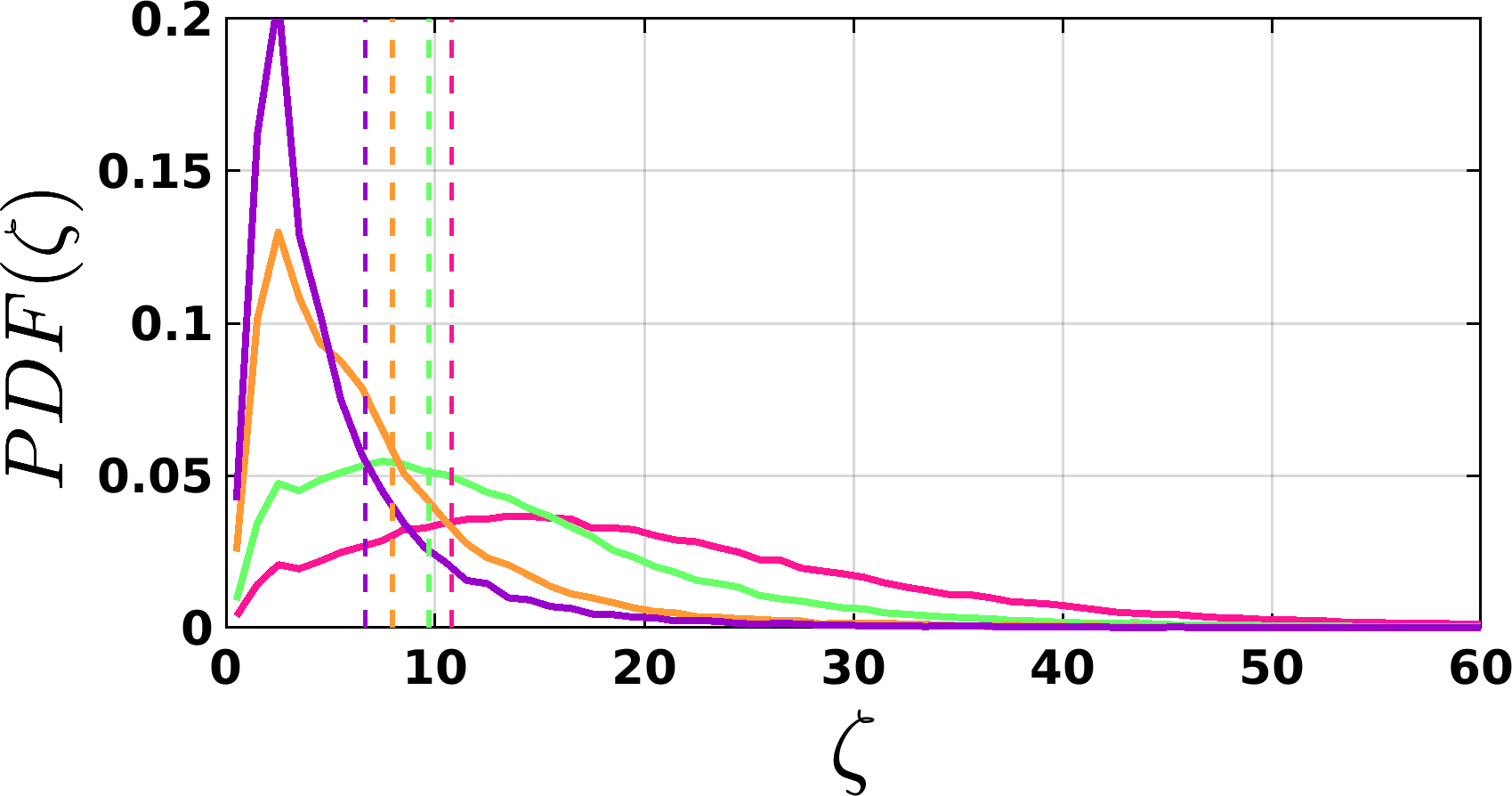}\\
\includegraphics[angle=0,width=0.5\textwidth]{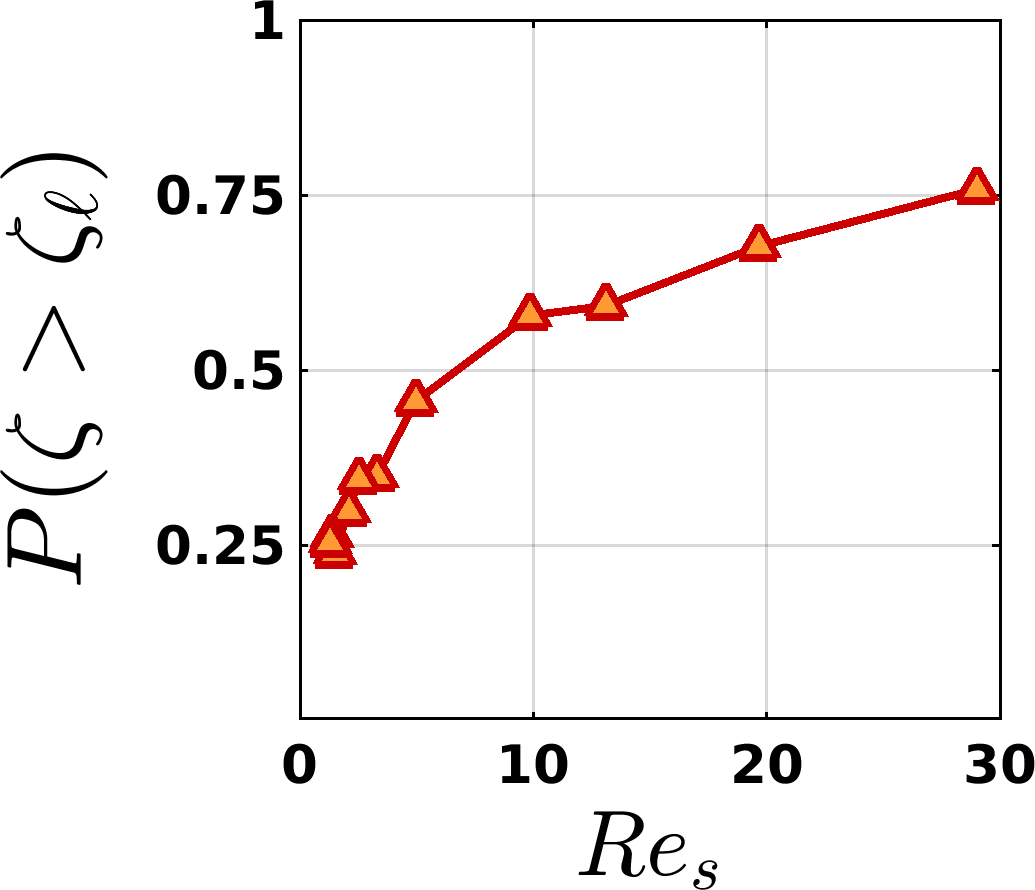}
\caption{
(Top): $PDF(\zeta)$ for $Re_s= 1.25, \, 2.5, \, 9.8, \, 29$ (violet, orange, light green and magenta solid lines, respectively) 
for the set of simulations where the radius of the injected rings $R=D/2$. 
The vertical dashed lines correspond to $\zeta_\ell=1/\ell$ for each $Re_s$; (Bottom): Probability $P(\zeta > \zeta_\ell)$ of observing structures
at length scales smaller than $\ell$ as a function of the superfluid Reynolds number $Re_s$. 
}
\label{fig:R1}
\end{figure}

\newpage

\section{Temporal evolution of curvature}\label{appx:E}

\begin{figure}[ht]
\centering
\begin{minipage}[c]{0.7\columnwidth}
\includegraphics[angle=0,width=1\textwidth]{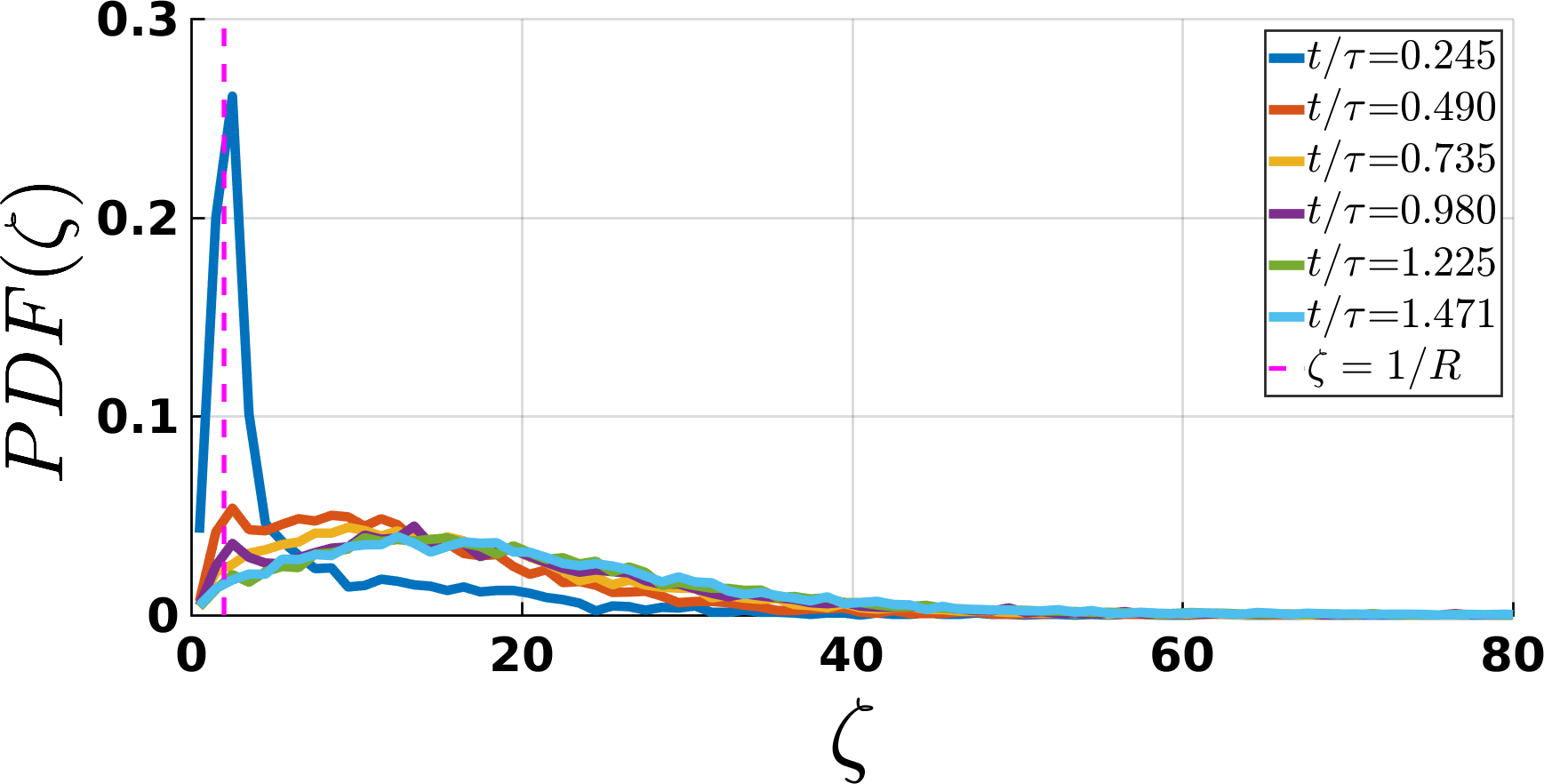}
\end{minipage}\\
\vspace{0.10\columnwidth}
\begin{minipage}[c]{0.7\columnwidth}
\includegraphics[angle=0,width=1\textwidth]{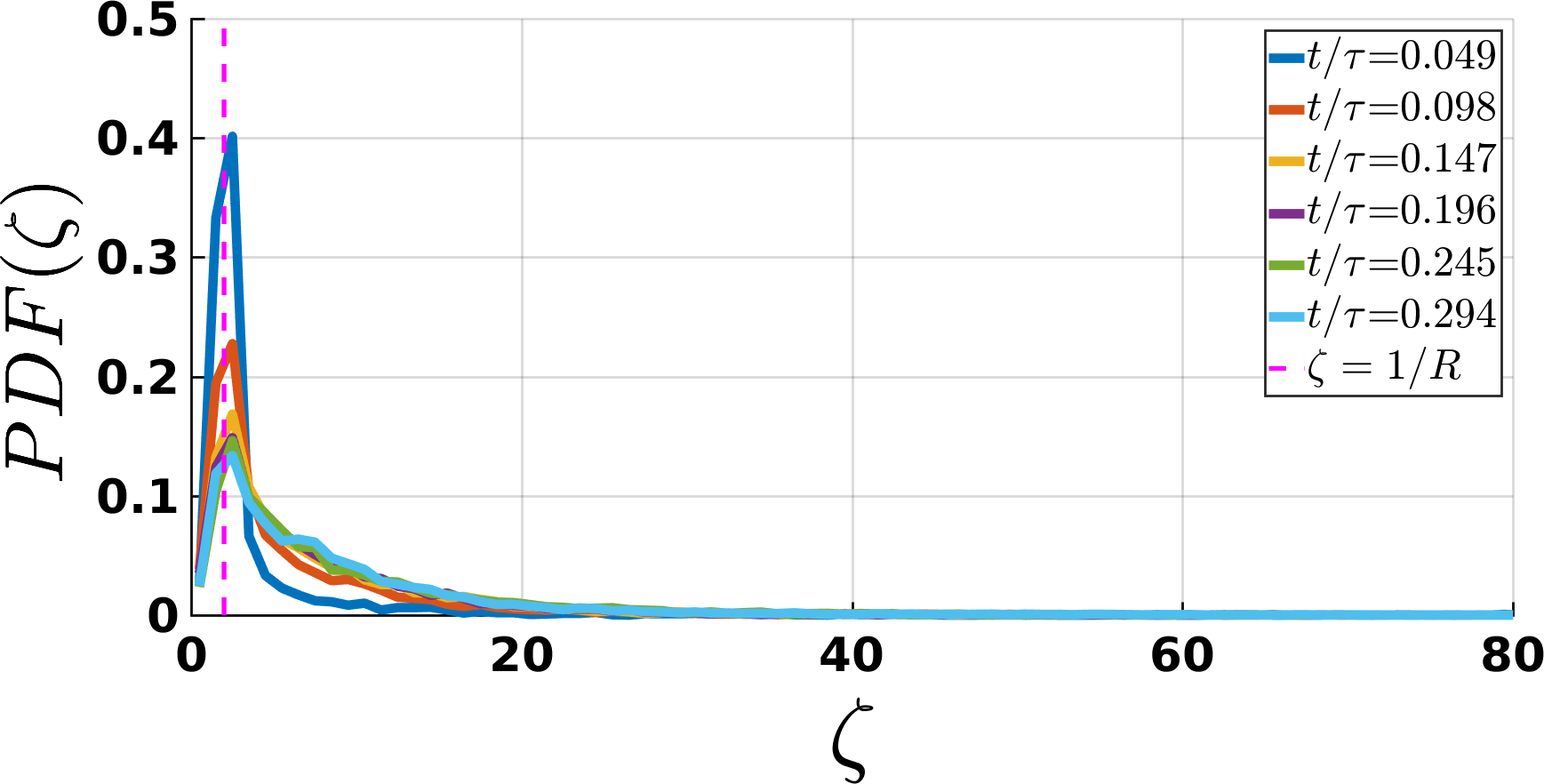}
\end{minipage}
\caption{Temporal evolution of the PDF of the curvature $\zeta$ (in $cm^{-1}$) 
for $Re_s=29$ and ${\dot L}_{inj}=3.35 \rm cm^{-2}  \rm s^{-1}$ (top) 
and $Re_s=1.25$ and ${\dot L}_{inj}=22.50 \rm cm^{-2}  \rm s^{-1}$ (bottom). In both simulations,
at saturation the vortex line density $\overline{L}$ is approximately equal to $120 \rm cm^{-2}$ and 
the radius of the injected vortex rings is $R=D/2$. We clearly observe the generation of smaller scale structures when $Re_s$
is larger.
}
\label{fig:S4}
\end{figure}

In this Appendix \ref{appx:E}, we show the temporal evolution of the PDF of the curvature $\zeta$.
We focus on two simulations, whose vortex-tangle snapshots are illustrated in Fig.~1 (a) and (b).
In the first simulation, $Re_s=29$ and ${\dot L}_{inj}=3.35 \rm cm^{-2}  \rm s^{-1}$ 
(corresponding temporal evolution of PDF($\zeta$) reported in Fig.~\ref{fig:S4} (top)), in the second
$Re_s=1.25$ and ${\dot L}_{inj}=22.50 \rm cm^{-2}  \rm s^{-1}$ (PDFs showed in Fig.~\ref{fig:S4} (bottom)). At saturation, 
in both numerical simulations the vortex line density is approximately equal to $120 \rm cm^{-2}$.

In both simulations, the radius of the injected vortex rings is $R=D/2$ and the corresponding curvature $\zeta=1/R=2/D$
is indicated in Fig.~\ref{fig:S4} (top) and (bottom) by a magenta dashed vertical line. The pattern which emerges from Fig.~\ref{fig:S4}
is clear: as the rings are injected, by interacting and reconnecting with the pre-existing tangle, smaller structures with corresponding
higher curvatures are generated. As $Re_s$ is increased (or, equivalently, as temperature is decreased), the smaller value
of the friction coefficients allows the generation and the survival of smaller structures with higher values of curvature: the resulting
PDF($\zeta$) is more shifted to the right. It is this generation of smaller scale structures as $Re_s$ increases which is responsible
for the observed plateau of the normalized kinetic energy dissipation $\widetilde{\epsilon}$ at large $Re_s$, 
reported in Fig.~\ref{fig:3} (a) and which is the principal finding of our work.

\newpage

\section{Alternative normalisation for Vinen turbulence}\label{appx:D}

\red{
In Fig.~\ref{fig:3} (a), the energy dissipation rates of all data sets are
normalised by the traditional factor $\langle I \rangle/\langle U \rangle^3$ used
in classical turbulence. This is because the main result (the red curve) refers to
a regime of quantum turbulence with the classical property that there is
an inertial range where a dissipationless cascade takes place. 
}

\begin{figure}[!htbp]
\centering
\includegraphics[angle=0,width=0.6\textwidth]{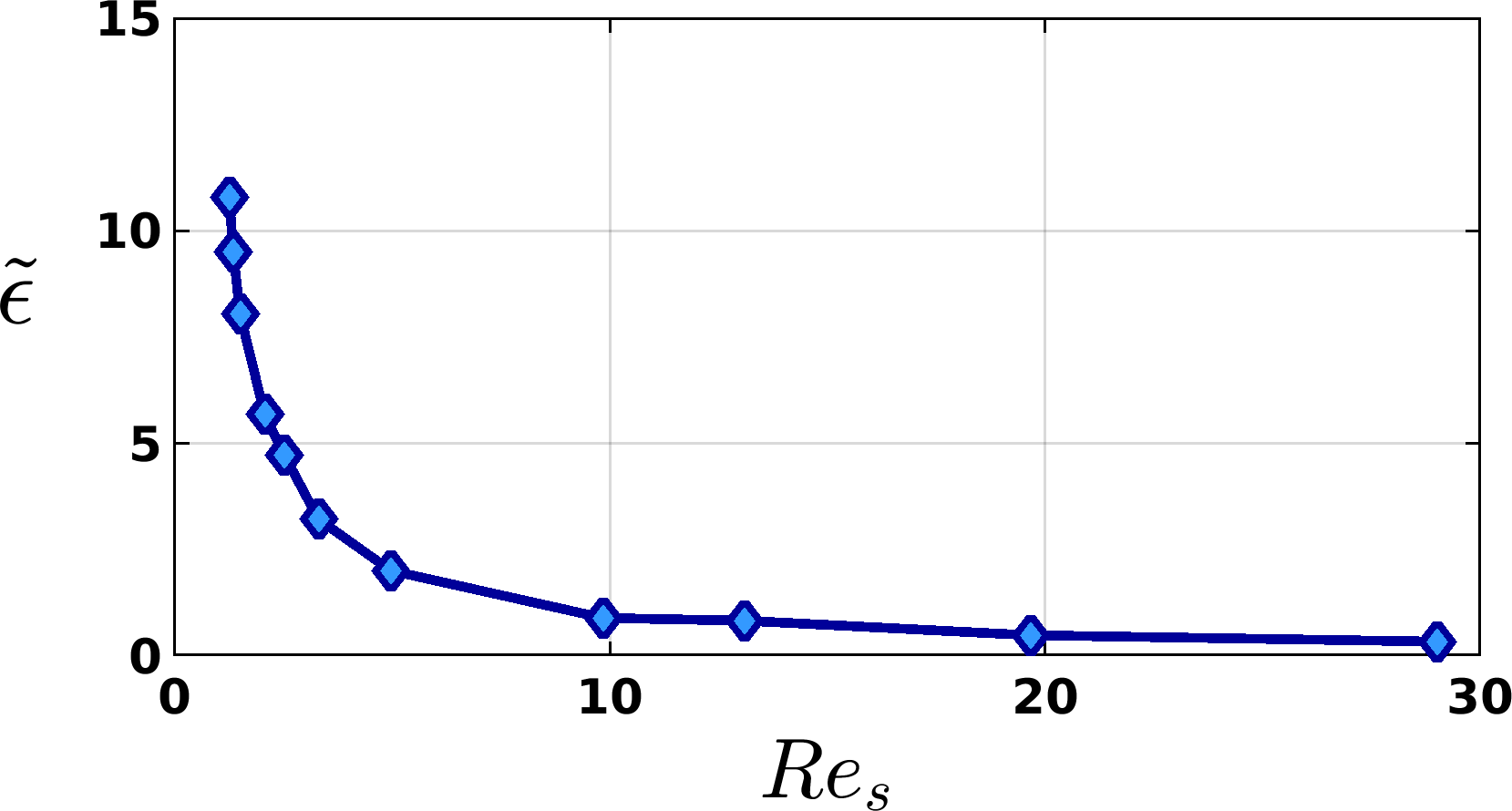}\\
\caption{
\red{
Time-averaged energy dissipation $\epsilon$ for Vinen turbulence obtained
by injecting vortex rings of radius
$R=\ell/2$, corresponding to the blue curve of Fig~\ref{fig:3}(a),
normalised by $\epsilon_{\ell}=v_s(\ell)^3/\ell \sim \kappa^3 \overline{L}^2$.
}
}
\label{fig:RX}
\end{figure}

\red{
Fig.~\ref{fig:3} (a) shows also data (green and blue curves) which refer to regimes of
Vinen-like quantum turbulence: the same flattening of $\widetilde{\epsilon}$ at
increasing $Re_s$ is apparent because the analogy
to the classical dissipation anomaly is independent of the
dynamics at the large scales of the flow. It is however natural to ask how the curves 
would look using an alternative normalisation. A dedicated normalisation factor for Vinen
turbulence would be the characteristic dissipation rate at scales comparable to the 
inter-vortex distance $\ell$, given by 
$\epsilon_{\ell}=v_s(\ell)^3/\ell \sim \kappa^3 \overline{L}^2$.
Using this normalisation factor, the resulting dissipation rate still resembles
the classical counterpart, as shown for example in Fig.~\ref{fig:RX}.
}

\newpage

\section{Numerical resolution of the small length scales}\label{appx:B}

\begin{figure}[!htbp]
\begin{minipage}[c]{0.40\columnwidth}
\includegraphics[angle=0,width=1\textwidth]{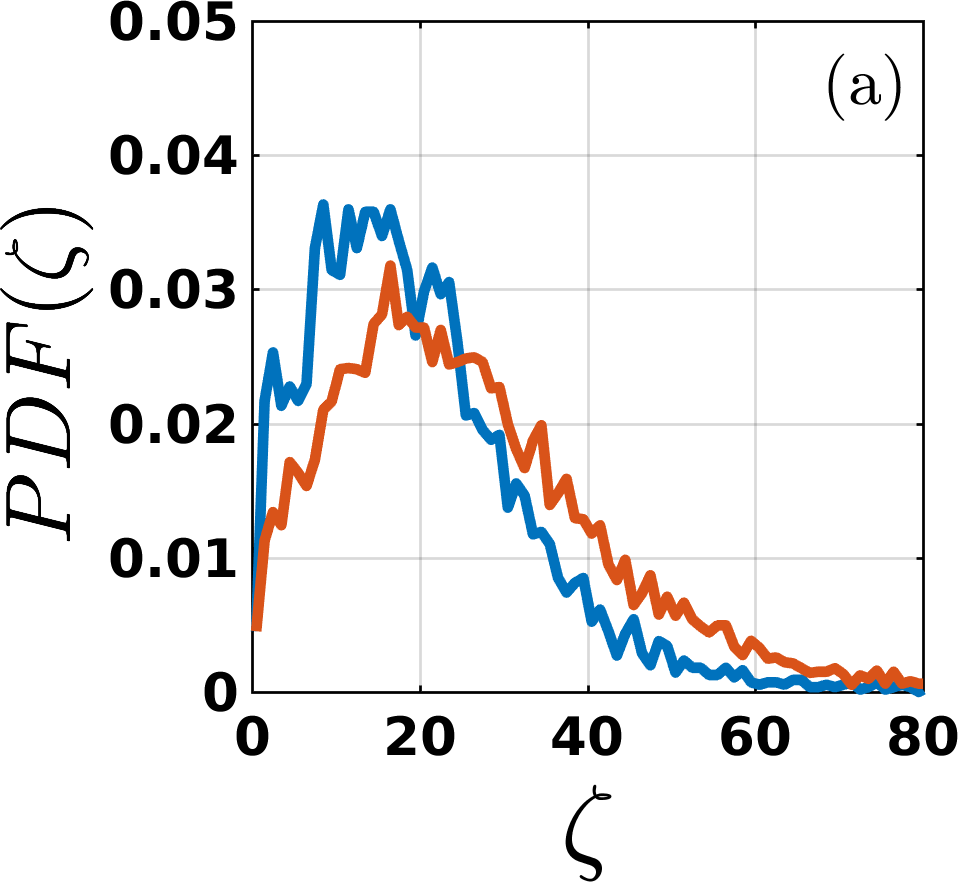}
\end{minipage}
\hspace{0.1\columnwidth}
\begin{minipage}[c]{0.4\columnwidth}
\includegraphics[angle=0,width=1\textwidth]{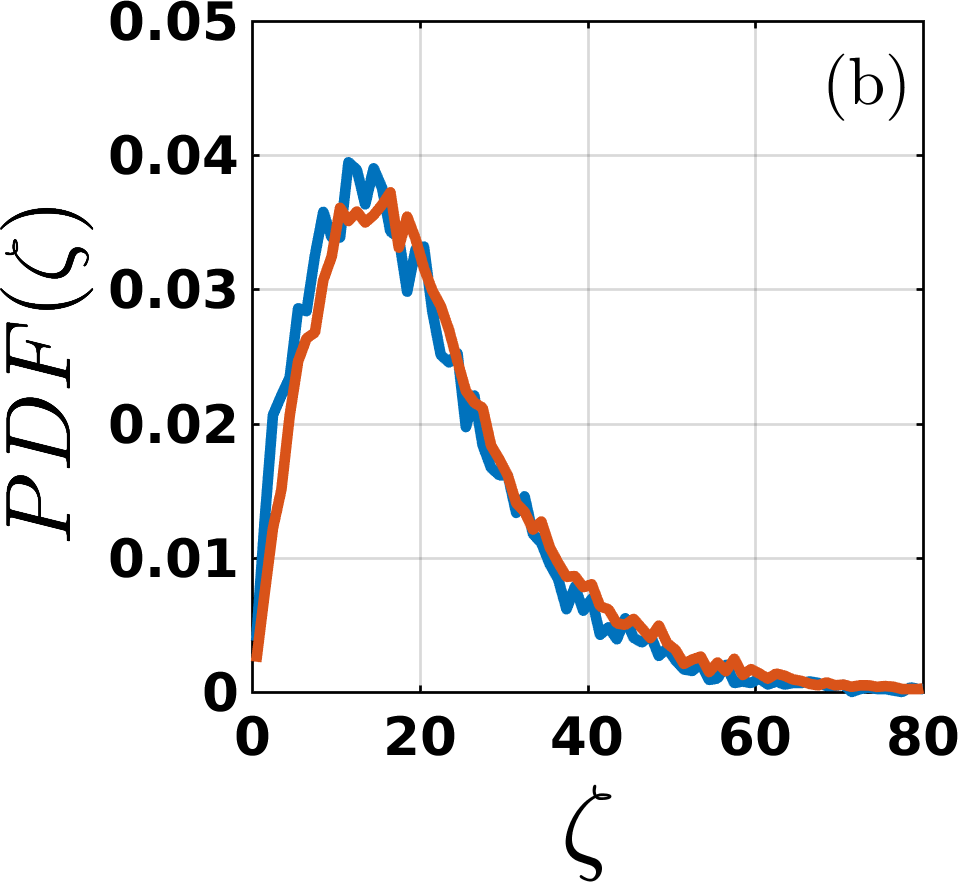}
\end{minipage}\\
\begin{minipage}[c]{0.40\columnwidth}
\includegraphics[angle=0,width=1\textwidth]{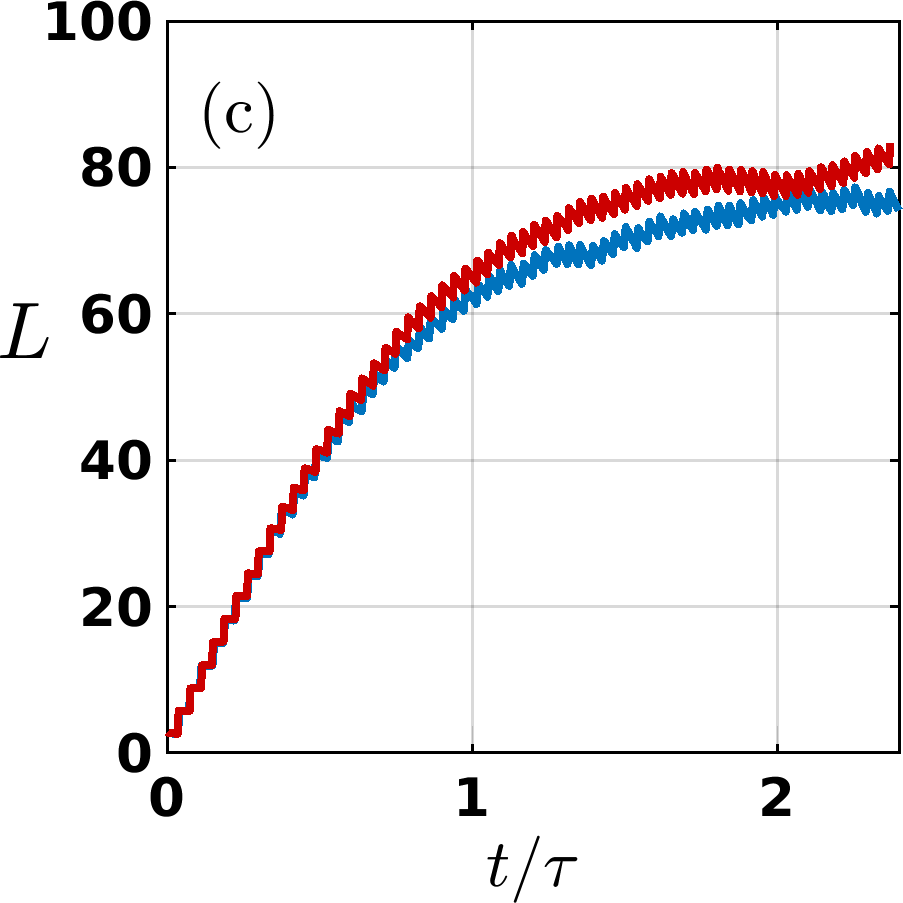}
\end{minipage}
\hspace{0.1\columnwidth}
\begin{minipage}[c]{0.40\columnwidth}
\includegraphics[angle=0,width=1\textwidth]{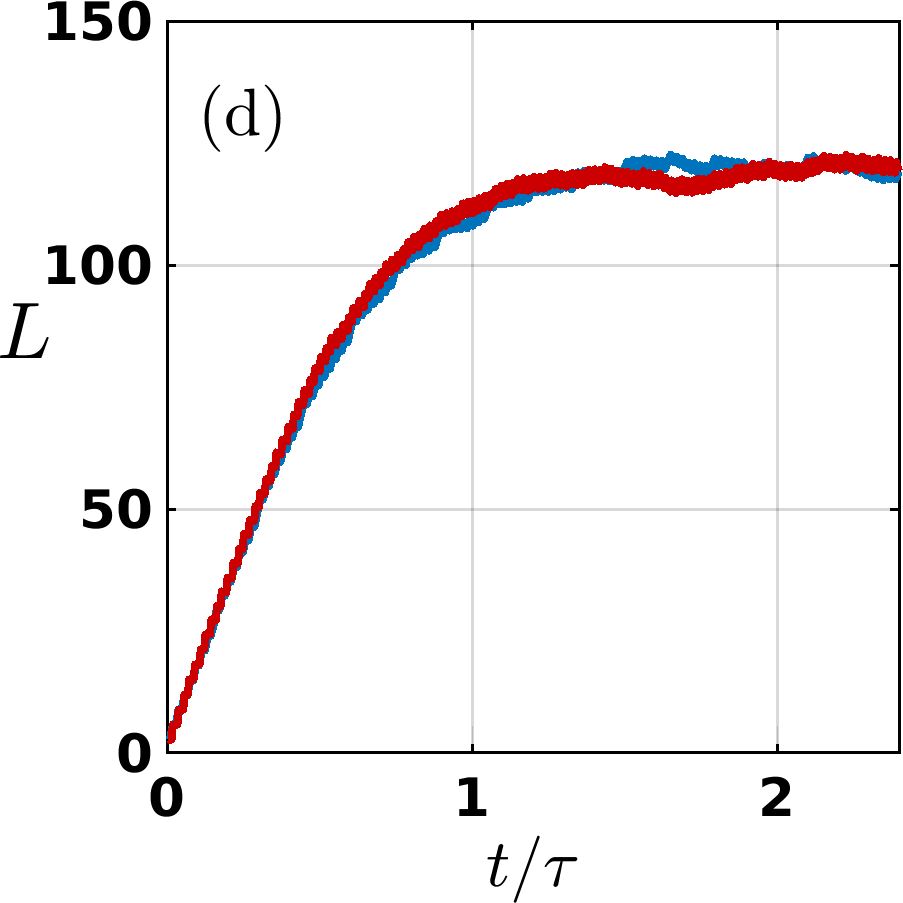}
\end{minipage}
\caption{Left (right) column refers to simulation where $Re_s=49.45$ and $\dot{L}_{inj}=1.0 \rm cm^{-2}\rm s^{-1}$
($Re_s=29$ and $\dot{L}_{inj}=3.35 \rm cm^{-2}\rm s^{-1}$). Blue (red) curves refer to spatial discretization $\delta=0.02$ cm
($\delta=0.01$ cm).
(a) and (b): $\rm{PDF}(\zeta)$ (in $\rm cm$) \textit{vs} curvature $\zeta$ (in $\rm cm^{-1}$);
(c) and (d): vortex-line density $L$ (in $\rm cm^{-2}$) \textit{vs} rescaled time $t/\tau$, 
where $\tau=2\pi/(\kappa \bar{L})$, $\bar{L}$ being the vortex-line
density at saturation.}
\label{fig:S0}
\end{figure}

As increasing $Re_s$ excites smaller length scales along the vortex lines 
it is natural to ask whether our numerical discretization correctly 
resolves these small length scales.  
To assess our numerical resolution we have repeated all the
simulations replacing $\delta$ with $\delta/2$. In the
analysis which leads to the calculation of $\widetilde{\epsilon}$ (Fig.~ \ref{fig:3}(a)), we have
rejected the results of the simulations, identified by the pair 
$(Re_s,\dot{L}_{inj})$, which do not satisfy either of 
the following strict criteria:
(i) the saturation value $\overline{L}$ obtained using
the two numerical resolutions are within $8\%$ of each other, 
and (ii) the PDFs of the curvature $\zeta$ overlap. 
The first criterion ensures that the turbulent intensity is correctly
captured, while the second is necessary in order to resolve accurately the curvature,
which governs the dissipation (Eq.~\ref{eq:epsilon_ring}). 
For example, Fig.~\ref{fig:S0} (b)
shows that for ($Re_s=29,\dot{L}_{inj}=3.35 \rm cm^{-2}\rm s^{-1}$)
the PDFs of the curvature do indeed overlap, 
hence criterion (ii) is satisfied, while this is not the case
for ($Re_s=49.45,\dot{L}_{inj}=1.0 \rm cm^{-2}\rm s^{-1}$),
see Fig.~\ref{fig:S0} (a). In Fig.~\ref{fig:S0} (c) and (d) 
we report the correspondent temporal evolution of the vortex-line density,
showing the impact of the resolution on this integral quantity:
in terms of criterion (i) simulation ($Re_s=49.45,\dot{L}_{inj}=1.0 \rm cm^{-2}\rm s^{-1}$)
lacks of spatial resolution along the vortex-lines, while simulation 
($Re_s=29,\dot{L}_{inj}=3.35 \rm cm^{-2}\rm s^{-1}$) is sufficiently resolved.

We remark
that this is the first time that such as strict criterion has been used
to test the VFM at low
temperatures or at high vortex line density; all previous
work was mainly concerned with properties at the large length scales,
whereas here we are primarily concerned with the smaller dissipation 
length scales. 
In practice, our strict criterion limits us to temperatures above 
$T \approx 1.3~\rm K$, above \cite{vinen-niemela-2002,vinen-2005,walmsley-etal-2014b} the appearance of scaling behaviour
for the KW cascade \cite{KozikSvistunov,LvovNazarenko,Krstulovic},
which, in the absence of dissipation, would shift energy to 
length scales of the order of $a_0$, not computationally resolvable by the VFM.
At such short scales acoustic emission \cite{GalantucciNature2021} and 
excitations of Carol-Matricon states dissipate the turbulent kinetic energy.


\newpage
\begin{thebibliography}{10}

\bibitem{Sreeni1984}
K.R. Sreenivasan,
On the scaling of the turbulence energy dissipation rate,
Phys. Fluids {\bf 27}, 1048 (1984).

\bibitem{KanedaIshihara2003}
Y. Kaneda, T. Ishihara, M. Yokokawa, K. Itakura and A. Uno,
Energy dissipation rate and energy spectrum in high resolution direct
numerical simulations of turbulence in a periodic box.
Phys. Fluids {\bf 15} L21 (2003)

\bibitem{Onsager1949}
L. Onsager, Statistical hydrodynamics,
Nuovo Cimento, Suppl. {\bf 6}, 279 (1949).

\bibitem{EyinkSreeni2006}
G.L. Eyink and K.R. Sreenivasan,
Onsager and the theory of hydrodynamic turbulence,
Rev. Modern Phys. {\bf 78}, 87 (2006).

\bibitem{BLR2014}
C.F. Barenghi, V. L'vov, and P.-E. Roche,
Experimental, numerical and analytical velocity spectra in
turbulent quantum fluid,
Proc. Nat. Acad. Sci. USA, {\bf 111} (suppl. 1) 4683 (2014).

\bibitem{stalp-skrbek-donnelly-1999}
S. R. Stakp, L. Skrbek and R. J. Donnelly,
Decay of Grid Turbulence in a Finite Channel,
Phys Rev Lett {\bf 82}, 4831 (1999)

\bibitem{maurer-tabeling-1998}
J. Maurer and P. Tabeling,
Local investigation of superfluid turbulence,
Europhys Lett {\bf 43}, 29 (1998).

\bibitem{salort-etal-2010}
J. Salort, C. Baudet, B. Castaing, B. Chabaud, F. Daviaud, T. Didelot, P. Diribarne, 
B. Dubrulle, Y. Gagne, F. Gauthier, A. Girard, B Henbral, B. Rousset, P. Thibault, P.E. Roche,
Phys Fluids {\bf 22}, 125102 (2020)

\bibitem{kolmogorov-1941}
A.N. Kolmogorov,
The {L}ocal {S}tructure of {T}urbulence in an {I}ncompressible {V}iscous {F}luid for {V}ery {L}arge {R}eynolds {N}umbers,
Dokl. Akad. Nauk. SSSR {\bf 30}, 301 (1941)

\bibitem{friction}
\red{
C.F. Barenghi, W.F. Vinen and R.J. Donnelly,
Friction on quantized vortices in helium~II: a review,
J. Low Temp. Phys. {\bf 52}, 189 (1982).
}

\bibitem{Nore1997}
\red{
C. Nore, M. Abid and M.E. Brachet,
Kolmogorov turbulence in low-temperature superflows,
Phys. Rev. Lett. {\bf 78}, 3896 (1997).
}

\bibitem{HillsRoberts}
R.N. Hills \& P.H. Roberts, Superfluid mechanics for a high density of
vortex lines, Arch. Rat. Mech. Anal. {\bf 66}, 43 (1977).


\bibitem{Finne}
A.P. Finne, T.Araki, R. Blaauwgeers, V.B. Eltsov, N.B. Kopnin,
M. Krusius, L. Skrbek, M. Tsubota \& G.E. Volovik,
An intrinsic velocity-independent criterion for superfluid turbulence,
Nature {\bf 424}, 1022 (2003).

\bibitem{Schwarz}
K.W. Schwarz,
Three-dimensional vortex dynamics in superfluid $^4$He:
homogeneous superfluid turbulence,
Phys. Rev. B {\bf 38}, 2398 (1988).

\bibitem{HanninenBaggaley}
R. H\"anninen and A.W. Baggaley,
Vortex filament method as a tool for computational
visualization of quantum turbulence,
Proc. Nat. Acad. Sci. USA {\bf 111} (suppl. 1), 4667 (2014).

\bibitem{schwarz-1985}
K. W. Schwarz,
Three-dimensional vortex dynamics in superfluid He 4: Line-line and line-boundary interactions
Phys. Rev. B {\bf 31}, 5782 (1985).

\bibitem{Kivotides2000}
D. Kivotides, C.F. Barenghi, and D.C. Samuels,
Triple vortex ring structure in superfluid helium II,
Science {\bf 290} 777 (2000).

\bibitem{Galantucci2020}
L. Galantucci, A.W. Baggaley, C.F. Barenghi, and G. Krstulovic,
A new self-consistent approach of quantum turbulence in superfluid helium
Eur. Phys. J. Plus {\bf 135}, 1 (2020).

\bibitem{Galantucci2022}
L. Galantucci, G. Krstulovic, and C.F. Barenghi,
Friction-enhanced lifetime of bundled quantum vortices,
arXiv:2107.07768 (2022)

\bibitem{Kivotides2007}
D. Kivotides,
Relaxation of superfluid vortex bundles via energy transfer to the
normal fluid,
Phys. Rev. B {\bf 76}, 054503 (2007).

\bibitem{Morris2008}
K. Morris, J, Koplik, and D.W.L. Rousen,
Vortex locking in direct numerical simulations of quantum turbulence,
Phys. Rev. Lett. {\bf 101}, 015301 (2008).

\bibitem{Walmsley2008}
P. M. Walmsley and A. I. Golov,
Quantum and quasiclassical types of superfluid turbulence,
Phys. Rev. Lett. {\bf 100}, 245301 (2008).

\bibitem{Leonard1985}
A. Leonard,
Computing three dimensional incompressible flows with votrex elements,
Ann. Rev. Fluid Mech. {\bf 17}, 523 (1985).

\bibitem{Yurkina2021}
O. Yurkina and S. K. Nemirovskii,
On the energy spectrum of the 3D velocity
field, generated by an ensemble of vortex loops,
Low Temp. Phys. {\bf 47}, 652 (2021).

\bibitem{walmsley-etal-2014b}
P. M. Walmsley,  D.E. Zmeev, F. Pakpour, and A.I. Golov,
Dynamics of quantum turbulence of different spectra,
Proc Natl Acad Sci USA {\bf 111}, 4691 (2014)

\bibitem{SkrbekSreeni}
L. Skrbek and K.R. Sreenivasan,
Developed quantum turbulence and its decay,
Phys. Fluids {\bf 24}, 011301 (2012).


\bibitem{vinen-2005}
W. F. Vinen,
Theory of quantum grid turbulence in superfluid $^3$He-B,
Phys. Rev. B {\bf 71}, 024513 (2005).

\bibitem{lvov-nazarenko-volovik-2004}
V. S. L'vov, S. V. Nazarenko and G. E. Volovik
Energy Spectra of Developed Superfluid Turbulence
JETP Letters {\bf 80}, 479 (2004)


\bibitem{Skrbek2021}
L. Skrbek,
Phenomenology of quantum turbulence in
superfluid helium,
Proc. Natl. Acad. Sci. USA {\bf 116}, e2018406118 (2021).

\bibitem{KozikSvistunov}
E.V. Kozik and B.V. Svistunov, 
Kelvin-wave cascade and decay of superfluid turbulence,
Phys. Rev. Lett. {\bf 92}, 035301 (2004).

\bibitem{LvovNazarenko}
V.S. L’vov and S.V. Nazarenko, 
Spectrum of Kelvin-wave turbulence in superfluids, 
JETP Lett. {\bf 91}, 428 (2010).

\bibitem{Krstulovic}
G. Krstulovic,
Kelvin-wave cascade and dissipation in low-temperature superfluid vortices
Phys. Rev. E {\bf 86}, 055301(R) (2012).

\bibitem{vinen-niemela-2002}
W. F. Vinen and J. J. Niemela,
Quantum Turbulence,
J Low Temp. Phys. {\bf 128}, 167 (2002).

\bibitem{vinen-1957c}
W. F. Vinen,
Mutual Friction in a heat current in liquid helium II. III. Theory of mutual friction,
Proc. R. Soc. London A {\bf 242}, 493 (1957).






\bibitem{gao-etal-2018}
J. Gao, W. Guo, S. Yui, M. Tsubota, W. F. Vinen,
Dissipation in quantum turbulence in superfluid $^4$He above 1 K
Phys. Rev. B {\bf 97}, 184518 (2018).

\bibitem{BDV1983}
C. F. Barenghi, R. J. Donnelly and W. F. Vinen,
Friction on Quantized Vortices in Helium II. A Review
J. Low Temp. Phys. {\bf 52}, 189 (1983).

\bibitem{Baggaley2012}
A. W. Baggaley, C. F. Barenghi, and Y. A. Sergeev,
Quasiclassical and ultraquantum decay of superfluid turbulence,
Phys. Rev. B {\bf 85}, 060501(R) (2012)

\bibitem{Araki2002}
\red{
T. Araki, M. Tsubota, and S.K. Nemirovskii,
Energy spectrum of superfluid turbulence with no normal-fluid component,
Phys. Rev. Lett. {\bf 89}, 145301 (2002).
}


\bibitem{makinen-etal-2022}
J.T. Makinen, S. Autti, P.J. Heikkinen, J.J. Hosio, R. Hänninen,
V.S. L’vov, P.M. Walmsley, V.V. Zavjalov, and V.B. Eltsov 
Rotating quantum wave turbulence
arXiv:2203.11527 (2022)

\bibitem{john-etal-2021}
J.P. John, D.A. Donzis, and K.R. Sreenivasan,
Does dissipative anomaly hold for compressible turbulence?,
J. Fluid Mech. {\bf 920}, A20 (2021)

\bibitem{DB}
R.J. Donnelly and C.F. Barenghi,
The observed properties of liquid helium at saturated vapour
pressure,
J. Phys. Chem. Ref. Data {\bf 27}, 1217 (1998)

\bibitem{GalantucciNature2021}
W. J. Kwon, G.  Del Pace, K. Xhani, L. Galantucci, A. Muzi Falconi, M. Inguscio,
F. Scazza and G. Roati
Sound emission and annihilations in a programmable quantum vortex collider.
Nature {\bf 600}, 64 (2021).


\end{thebibliography}
\end{document}